\documentclass[twocolumn]{aastex62}
 \def\mso{\,\mathrm{M}_\odot}

 \def\zso{\,{\rm Z}_\odot}
 \def\kms{\, {\rm km}\, {\rm s}^{-1}}

 \def\simle{\mathrel{\hbox{\rlap{\hbox{\lower4pt\hbox{$\sim$}}}\hbox{$<$}}}}
 \def\simgr{\mathrel{\hbox{\rlap{\hbox{\lower4pt\hbox{$\sim$}}}\hbox{$>$}}}}

\graphicspath{{./}{figures/}}

\received{06/02/2020}
\revised{07/03/2020}
\accepted{08/19/2020}

\submitjournal{ApJ}
\shorttitle{Pre-collapse Properties of SLSNe and lGRB Progenitor Models}
\shortauthors{Aguilera-Dena et al.}
\begin{document}

\title{Pre-collapse Properties of Superluminous Supernovae and Long Gamma-Ray Burst Progenitor Models}

\correspondingauthor{David R. Aguilera-Dena}
\email{davidrad@astro.uni-bonn.de}

\author[0000-0002-3874-2769]{David R. Aguilera-Dena}
\affiliation{Argelander-Institut f\"ur Astronomie, Universit\"at Bonn, Auf dem H\"ugel 71, 53121 Bonn, Germany}
\affiliation{Max-Planck-Institut f\"ur Radioastronomie, Auf dem H\"ugel 69, 53121 Bonn, Germany}

\author{Norbert Langer}
\affiliation{Argelander-Institut f\"ur Astronomie, Universit\"at Bonn, Auf dem H\"ugel 71, 53121 Bonn, Germany}
\affiliation{Max-Planck-Institut f\"ur Radioastronomie, Auf dem H\"ugel 69, 53121 Bonn, Germany}

\author[0000-0003-4453-3776]{John Antoniadis}
\affiliation{Max-Planck-Institut f\"ur Radioastronomie, Auf dem H\"ugel 69, 53121 Bonn, Germany}
\affiliation{Argelander-Institut f\"ur Astronomie, Universit\"at Bonn, Auf dem H\"ugel 71, 53121 Bonn, Germany}
\affiliation{Institute of Astrophysics, FORTH, 
Dept. of Physics, University of Crete,
Voutes, University Campus, GR-71003 Heraklion, Greece}

\author[0000-0002-4470-1277]{Bernhard M\"uller}
\affiliation{School of Physics and Astronomy, Monash University, VIC 3800, Australia}

\begin{abstract}

We analyze the properties of 42 rapidly rotating, low metallicity, quasi-chemically homogeneously evolving stellar models in the mass range between 4 and 45 $\mso$ at the time of core collapse. Such models were proposed as progenitors for both superluminous supernovae (SLSNe) and long duration gamma-ray bursts (lGRBs), and the Type Ic-BL supernovae (SNe)
 that are associated with them. Our findings suggest that whether these models produce a magnetar driven SLSN explosion or a near-critically rotating black hole (BH) is not a monotonic function of the initial mass. Rather, their explodability varies non-monotonically depending on the late core evolution, once chemical homogeneity is broken. Using different explodability criteria we find that our models have a clear preference to produce SLSNe at lower masses, and lGRBs at higher masses; but find several exceptions, expecting lGRBs to form from stars as low as 10 $\mso$, and SLSNe with progenitors as massive as 30 $\mso$. In general, our models reproduce the predicted angular momenta, ejecta masses and magnetic field strengths at core collapse inferred for SLSNe and lGRBs, and suggest significant interaction with their circumstellar medium, particularly for explosions with low ejecta mass.

\end{abstract}


\section{Introduction} \label{sec:intro}

Superluminous supernovae (SLSNe) and long duration gamma-ray bursts (lGRBs) are highly energetic transients  that have generated great 
interest in astrophysics due to their extreme properties. Despite some differences in their environments and distributions  \citep[e.g.][]{2015MNRAS.449..917L,2016MNRAS.458...84A}, there is accumulating observational and theoretical evidence alluding to a similar origin \citep[e.g.][]{2014ApJ...787..138L,2016A&A...593A.115J,2018MNRAS.475.2659M}. 

Type I SLSNe (henceforth referred to simply as SLSNe) are characterized by intrinsic luminosities 10 to 100 times larger than those of ``typical'' supernovae \citep[SNe; see][for recent reviews]{2012Sci...337..927G,2017hsn..book..431H,2018SSRv..214...59M,2019NatAs...3..697I}. Their spectra show no hydrogen features, and helium features are rarely found in late time spectra of some events \citep{2016MNRAS.458.3455M}, suggesting that the envelopes of their progenitors contain no hydrogen very little to no helium; i.e., they are bare carbon and oxygen cores \citep{2019ApJ...882..102G}. This suggests that their progenitors undergo a phase of intense mass loss and/or mixing, during which these elements are efficiently depleted in their envelopes. Helium would likely be detectable in the SN spectra if $\simgr $0.1 $\mso$ of it was present in the ejecta  \citep[however, this estimate is somewhat sensitive to the distribution of  $^{56}$Ni in the SN ejecta, and the actual composition of the helium-rich layer,][]{2012MNRAS.422...70H,2015MNRAS.453.2189D,2019ApJ...872..174Y}. These SNe are found to prefer low metallicity environments \citep{2016ApJ...830...13P,2016A&A...593A.115J,2018MNRAS.473.1258S}, and their high bolometric luminosities of SLSNe are not easily  explained by radioactive $^{56}$Ni decay, and require an additional source of energy.

Although the nature of SLSN progenitors is still not known, some insight has been gained by studying  the several mechanisms that have been proposed to explain their observational properties. Some of the leading theories invoke a continuous energy deposition into the ejecta by the spin-down of a newly-formed millisecond magnetar (e.g. \citealt{2010ApJ...717..245K,2010ApJ...719L.204W,2015MNRAS.454.3311M,2017ApJ...850...55N}), the 
accretion of fallback material onto a central compact object \citep[e.g.][]{2018arXiv180600090M},  
interaction between the SN ejecta and the circumstellar medium \citep[CSM; e.g.][]{2012ApJ...746..121C}, or the formation of  large amounts of $^{56}$Ni (more than 3\,$\mso$), resulting from a pair-instability-driven explosion in a very massive star (e.g. \citealt{2009Natur.462..624G}).

In the context  of the magnetar model, which is considered the most likely explanation, 
a large sample of SLSN light curves have been analysed, setting constrains on the distribution of ejecta masses, magnetar spin periods and magnetic field strengths that are required to reproduce the observations \citep{2017ApJ...850...55N,2018ApJ...869..166V,2018ApJ...865....9B,2019ApJ...872...90B,2020arXiv200209508B}. These studies have found that the ejecta masses of SLSNe range between 3.6 and 40 $\mso$, setting them significantly apart from type Ib/c SNe, and that the periods and magnetic fields of the millisecond pulsars that input energy into the ejecta are in the range of 1 to 8 ms and 0.3 to 10 $\times 10^{14}$ G, respectively. In this work, we 
present an evolutionary channel for progenitors of SLSNe, relying on efficient rotational mixing in massive, low-metallicity stars, that can reproduce many of the properties inferred from the SLSN light curve samples.

lGRBs on the other hand, are distinguished by beamed 
emission of energetic gamma-rays formed in collimated relativistic jets \citep[see, for example][for a review]{2009grb..book.....V}.
They are observed preferentially in low metallicity environments \citep[e.g.][]{2019A&A...623A..26P}, and
their host galaxies are similar to those hosting SLSNe. Furthermore, they are associated with a sub-class of hydrogen- and helium-deficient SNe known as Type Ic-BL \citep{2012grb..book..169H}. These SNe are less luminous than SLSNe, but  still have large ejecta velocities with kinetic energies that are typically 10 times larger than those of ``typical'' Type Ic SNe.

In the framework of the collapsar model \citep{1993ApJ...405..273W}, lGRBs form from
the collapse of a rapidly rotating stellar core into a black hole (BH). A corresponding
progenitor evolution is provided by rapidly rotating, rotationally mixed
models which undergo so called quasi-chemically homogeneous evolution \citep{1987A&A...178..159M,1992A&A...265L..17L,2000ApJ...544.1016H,2011A&A...530A.115B},
predicting lGRBs to occur in massive, rapidly rotating low metallicity single stars \citep{2005A&A...443..643Y,2006ApJ...637..914W,2006A&A...460..199Y} and in massive close binaries \citep{2008A&A...484..831D,2007A&A...465L..29C}.
Notably, homogeneously evolving massive close binaries may also explain massive BH
mergers (\citealt{2009A&A...497..243D}; \citealt{2016A&A...588A..50M}; \citealt{2020arXiv200711299H}) and ultra-luminous
X-ray sources \citep{2017A&A...604A..55M}.
Given that SLSNe and lGRBs present similarities in their 
environments, spectra, and energetics, \cite{2018ApJ...858..115A} (henceforth ALMS18) 
suggested a unified scenario for both types of transients, based on chemically homogeneous 
evolution, which leads to fast-rotating pre-collapse models with hydrogen- and helium-depleted envelopes. 

The mass loss and late evolution of these models is 
influenced significantly by intense neutrino cooling after core helium 
exhaustion. This accelerates the contraction of the core and enhances centrifugally-driven mass loss.  
 ALMS18 advocated that the end-product is either a ``collapsar''  and a fast-spinning BH \citep{1993ApJ...405..273W}, or a rapidly-spinning, highly-magnetized neutron star \citep[NS;][]{2010ApJ...717..245K,2010ApJ...719L.204W} --- giving rise to a lGRB or a SLSN respectively. It was suggested that lower mass progenitor models might correspond to SLSNe while more massive ones could correspond to lGRBs, but the analysis performed by \cite{2020arXiv200209508B} suggests that SLSN progenitors must have a wide range of final masses, from 3.6 to 40 $\mso$, in great contrast to progenitors of ``normal'' stripped-envelope SNe that have a strong cutoff in progenitor mass distribution above 10 $\mso$. We address the problem in this paper by studying the properties of such progenitor models at core-collapse.

While determining which stars will form
a NS or a BH is a highly complex astrophysical problem, 
several recent studies have established a series 
of simple diagnostic indicators to predict the final fate based on the  
pre-collapse stellar structure \citep[e.g.][]{2011ApJ...730...70O,2012ApJ...757...69U,2014ApJ...783...10S,2016ApJ...818..124E,2016MNRAS.460..742M,2018ApJ...860...93S}. One such diagnostic is the 
so-called core compactness parameter,  $\xi_{\text M}$ , which is 
motivated by  hydrodynamic simulations of neutrino-driven SN explosions \citep{2011ApJ...730...70O}. The behavior of this variable was found to 
be non-monotonic with initial mass, and depends mainly on the detailed
core behavior of an evolutionary sequence and the core mass it develops,
but is strongly affected by whether carbon burning proceeds radiatively 
or in a number of convective flames \citep{2019arXiv190500474S}, and more
weakly in other properties, such as the ratio of carbon to oxygen after helium burning, the chosen wind prescriptions, and the numerical 
treatment employed. Other proposed diagnostics of explodability  \citep[e.g][]{2016ApJ...818..124E,2016MNRAS.460..742M} generally yield 
similar results, i.e. they predict a transition around the dividing line between radiative and convective core carbon burning, with regimes where stars explode generally behaving non-monotonically.

Motivated by the findings of ALMS18, in this paper we explore the late burning stages and the progenitor-remnant connection for  
rapidly-rotating, chemically homogeneous massive stars. Expanding on 
our previous results, we evolve, for the first 
time, 42 rapidly-spinning stellar models to core 
collapse. We then apply the aforementioned 
diagnostic tools to predict their final fate and 
diagnose their relevance to lGRBs and 
magnetar-driven SLSNe. The text is organized as 
follows: In Section\,\ref{sec:method} we briefly 
describe the assumptions we employed to perform our 
calculations, and compare with the results of ALMS18. We present our analysis in Section\,\ref{sec:results}, highlighting that detailed core evolution during and after carbon burning will determine whether a chemically homogeneously evolving stellar model might result in a lGRB or a magnetar driven SLSN. In Section \ref{sec:obs} we infer observable properties of SLSNe from our models. We finalize this paper with a brief summary and conclusions in Section\,\ref{sec:conclusions}.

\section{Method}\label{sec:method}

The simulations presented in this work are similar to those performed in ALMS18. In this Section we review the methods and assumptions used, highlighting the similarities and differences to the previous simulations in Section \ref{sub:old}, and describe the consequences of these changes in Section \ref{sub:new}.

\subsection{Physical and numerical parameters}\label{sub:old}

We performed simulations of the evolution of massive single stars using the Modules for Experiments in Stellar Astrophysics code (MESA) in its version 10398 \citep{MESAI,MESAII,MESAIII,2018ApJS..234...34P}. 
To remain consistent to the simulations presented by ALMS18, our models have an initial 
equatorial rotation velocity of 600 $\kms$, and a metallicity of 
1/50$\zso$, scaled from solar metallicity \citep{1996ASPC...99..117G}. The initial value for the equatorial rotational velocity is chosen to represent the fastest rotating known O type star \citep{2013A&A...560A..29R}, and the combination of fast rotation and low metallicity are chosen to guarantee quasi-chemically homogeneous evolution. The choice of metallicity was initially conceived to be consistent with the results of \cite{2015A&A...581A..15S}, who modelled chemically homogeneously evolving massive stars to study the stellar population of the low metallicity galaxy I Zw 18. This value is smaller than the mean values observed for both SLSNe and lGRBs \citep[e.g.][]{2014ApJ...787..138L}, but chemically homogeneous evolution has been observed to be present at higher metallicities \citep[although all of them subsolar, e.g.][]{2011A&A...530A.115B}, so we expect that different parameter choices will produce similar progenitors, albeit probably over a smaller region of the parameter space, at such metallicities. Thus, the effect of varying metallicity and initial rotational velocity in the context of enhanced rotational mixing efficiency is beyond the scope of this paper.

The models cover the mass range  between 4 and 45$\mso$  
with a step size of 1$\mso$. Evolutionary sequences with an initial mass larger than 45$\mso$ were not computed since they are subject to pulsational pair instability near the core in their late burning stages. 
Likewise,  models with masses smaller than 4$\mso$ were excluded because
the initial rotational velocity applied was above their breakup velocity
at the zero-age main sequence (ZAMS).  Our simulations are carried out from ZAMS to core collapse.

These calculations have similar parameters as the Series\,B evolutionary sequences from ALMS18. They include rotational mixing, dominated by Eddington-Sweet circulation but also including secular and dynamical instabilities and the Goldreich--Schubert--Fricke instability, as described by \cite{2000ApJ...528..368H}. They evolve chemically homogeneously and have rotational mixing coefficients enhanced by a factor 10 compared to \cite{2011A&A...530A.115B}. This choice is made, similarly to ALMS18, to both decrease the amount of helium in the surface of our models at core collapse, and to extend the parameter space where chemically homogeneous evolution occurs into lower masses. While introduced as an ansatz, we justify this choice by the fact that these coefficients are not well parametrized in the case of near-critical rotation, as well as by the fact that, through this choice, we are able to reproduce features that are expected in SLSN and GRB progenitors. These include not only the rotation rates, enough to power the two energetic explosions, but the lower limit in ejecta masses of SLSNe found by  \cite{2020arXiv200209508B}, which is of the order of a few $\mso$, and the peculiar surface composition implied by the observed spectra.

Convection is applied according to the Ledoux criterion, and standard mixing length theory \citep{1958ZA.....46..108B} with $\alpha_{\text MLT} = 1.5$ is employed. Step convective overshooting during hydrogen burning with $\alpha_{\text OV} = 0.335$ and semiconvection with $\alpha_{\text SC} = 0.01$ are applied, and we employ MESA's \texttt{approx21} nuclear network \citep{MESAI}. We use Type 2 OPAL opacities to take into account the chemical enrichment of the surface \citep{1996ApJ...464..943I} and implement the empirical mass loss rates from \citep{2001A&A...369..574V} and \citep{1995A&A...299..151H} for hydrogen rich (X$_S>0.7$) and hydrogen poor stars (X$_S<0.4$), respectively; the latter multiplied by a factor of 1/10. In the regime in between we linearly interpolate between the two, and we employ a metallicity dependence of $\dot{\text{M}}\propto(\text{Z}/ \zso)^{0.85}$. Mass loss is, however, domina ted by the effect of rotation. We implement the enhancement of rotational mass loss rates due to rotation \citep{1993ApJ...409..429B}, and impose that the ratio $\Omega / \Omega_{\text{crit}}<0.98$, which has a larger effect than the prescriptions employed. Limiting $\Omega / \Omega_{\text{crit}}$ to be smaller than 1 has the effect of avoiding models that rotate above the critical limit, which has been observed in previous calculations of chemically homogeneoously evolving stars, that have used rotationally enhanced mass loss rates, but with an upper limit given by the thermal mass loss rate. The value of 0.98 is arbitrary, but varying it by a few percent was not observed to have an effect on our results.

Some changes were applied both to the numerical treatment and the employed physics to improve on the results of ALMS18 by reaching core collapse (infall velocity larger than 1000 km s$^{-1}$ in the core) in our evolutionary calculations, and to more accurately capture the effects of rotation. We include the treatment of the hydrodynamical structure of the star, but exclude the outer layers from this treatment when time steps are small (i.e. the velocity is set to 0 where $\log \left(\text{T}/\text{K}\right) < 8$ if the time steps are smaller than 0.1 yr; which corresponds to the envelope during the last phases of evolution). We neglect mass loss during the last years of evolution when the central temperature exceeds $2.5 \times 10^9 \ \text{K}$; and we include the predictive mixing scheme in MESA to avoid glitches in the structure of convective regions during helium burning. For models with initial mass between 40 and 45 $\mso$, we remove the condition that $\Omega / \Omega_{\text{crit}}<0.98$ at the end of carbon burning, which is only broken for a few timesteps and then recovered through rotationally enhanced mass loss.

We change the treatment of angular momentum loss, which has been modified in the most recent releases of MESA. Angular momentum loss was originally associated to the integrated amount of angular momentum contained in a layer of mass equivalent to the mass lost at each timestep. The current employed mechanism defines $\dot{J}$ as the product of the mass loss rate at each timestep and the specific angular momentum at the surface. This is equivalent to removing angular momentum from a layer closer to the surface, which has higher specific angular momentum than the average over a larger volume. This affects the final masses and compositions of these models with respect to the previously published ones. This corresponds to setting the variable do\_adjust\_J\_lost = .true. option, which is set by default in version 10398.

The uncertainties in the prescription of the specific angular momentum loss,
i.e., the angular momentum lost per lost gram of matter,
affect mostly the total amount of matter lost in the phases near critical rotation, since the required amount of angular momentum loss is simply dictated
by the decrease in the moment of inertia during the contraction stages.
An estimate of the inherent uncertainty can be obtained by comparing the results presented below with those of ALMS18 (cf. Section \ref{sub:new}).

The changes in the treatment of angular momentum losses likely represent an improvement in the calculations, but it is important to note that, even though angular momentum losses are treated in a more physically consistent way, the true angular momentum losses that a critically rotating star experiences are uncertain. The description of stellar structure is only accurate only up to a certain point, but assumptions that go into the calculations of rapidly rotating stars might break down above a certain threshold value of $v_{\text{rot}}/v_{\text{crit}}$, and the models presented here are certainly above that value as they are at critical rotation for a significant portion of their evolution.

Furthermore, fast rotation induces a latitude dependent temperature structure, which may lead to a latitude dependent mass and angular momentum loss rates, and the mass close to the surface of the star might interact with the magnetic field and the decretion disk, making this a highly non trivial problem. Therefore, although we have improved the physical treatment of angular momentum loss, our results are subject to the uncertainties that plague all rapidly rotating stellar evolution calculations, and should be taken with caution.

\subsection{Comparison with ALMS18}\label{sub:new}

Differences in the evolutionary sequences in ALMS18 and the current work (see Section \ref{sub:old}) are mainly due to the different treatment of the angular momentum loss. Having a more efficient angular momentum loss per unit mass implies that an almost critically rotating star will have to lose less mass as it contracts for its surface to remain under the break-up limit. Differences in angular momentum loss rate can be up to about 50\%, and they are especially significant during core helium burning, but remain important until a few decades before core collapse.

\begin{figure}[h!]
\epsscale{0.95}
\plotone{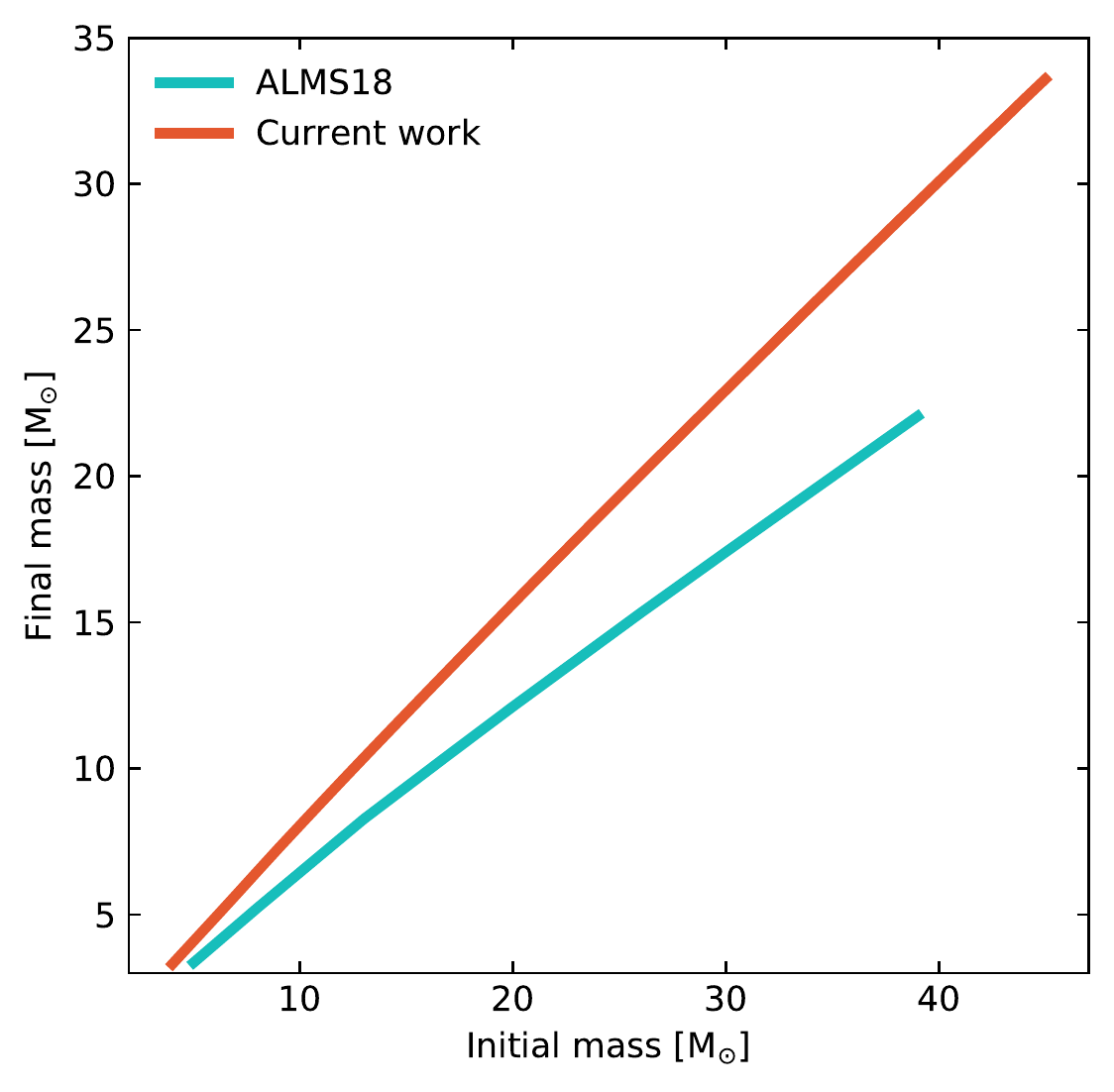}
\plotone{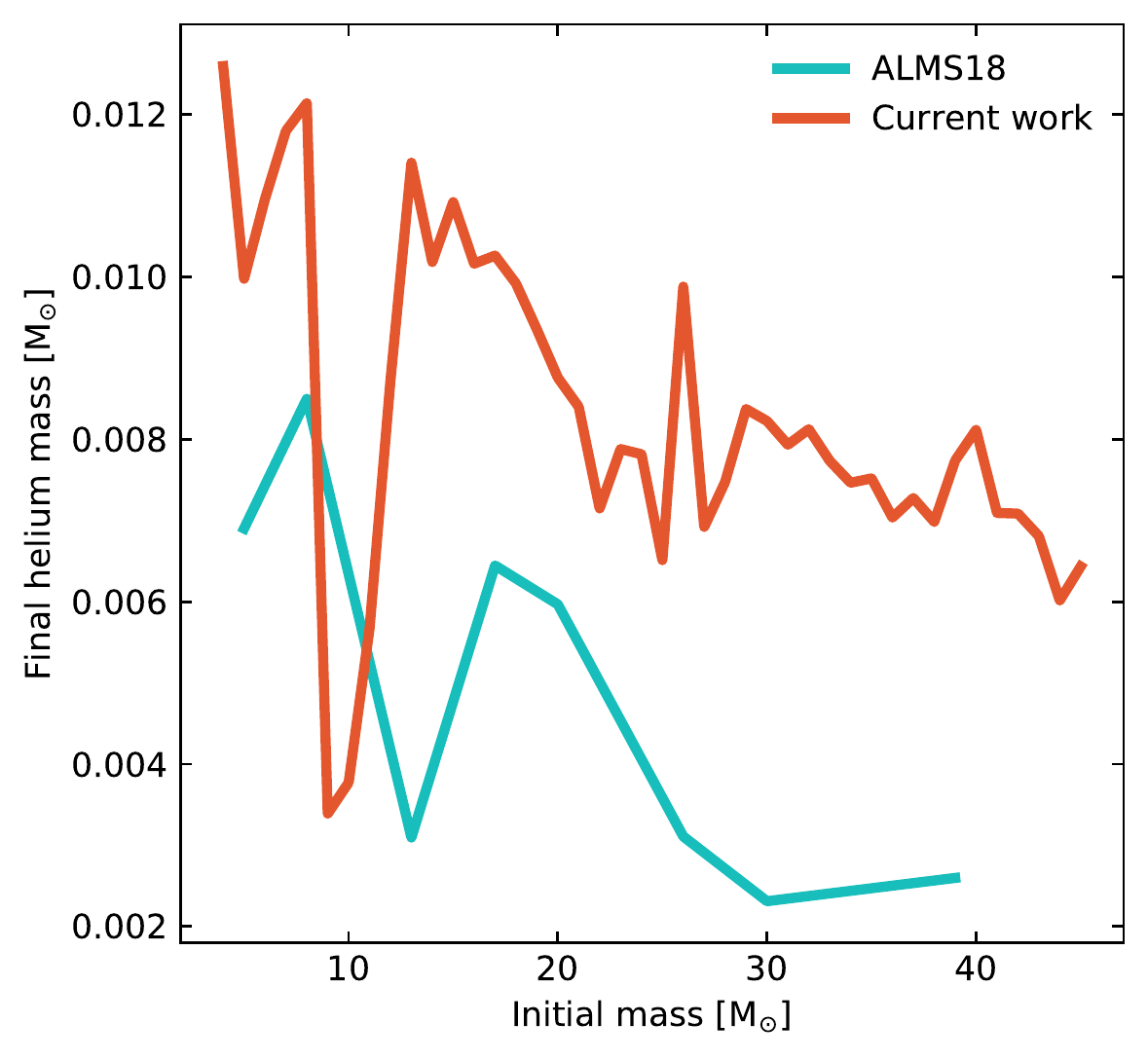}
\plotone{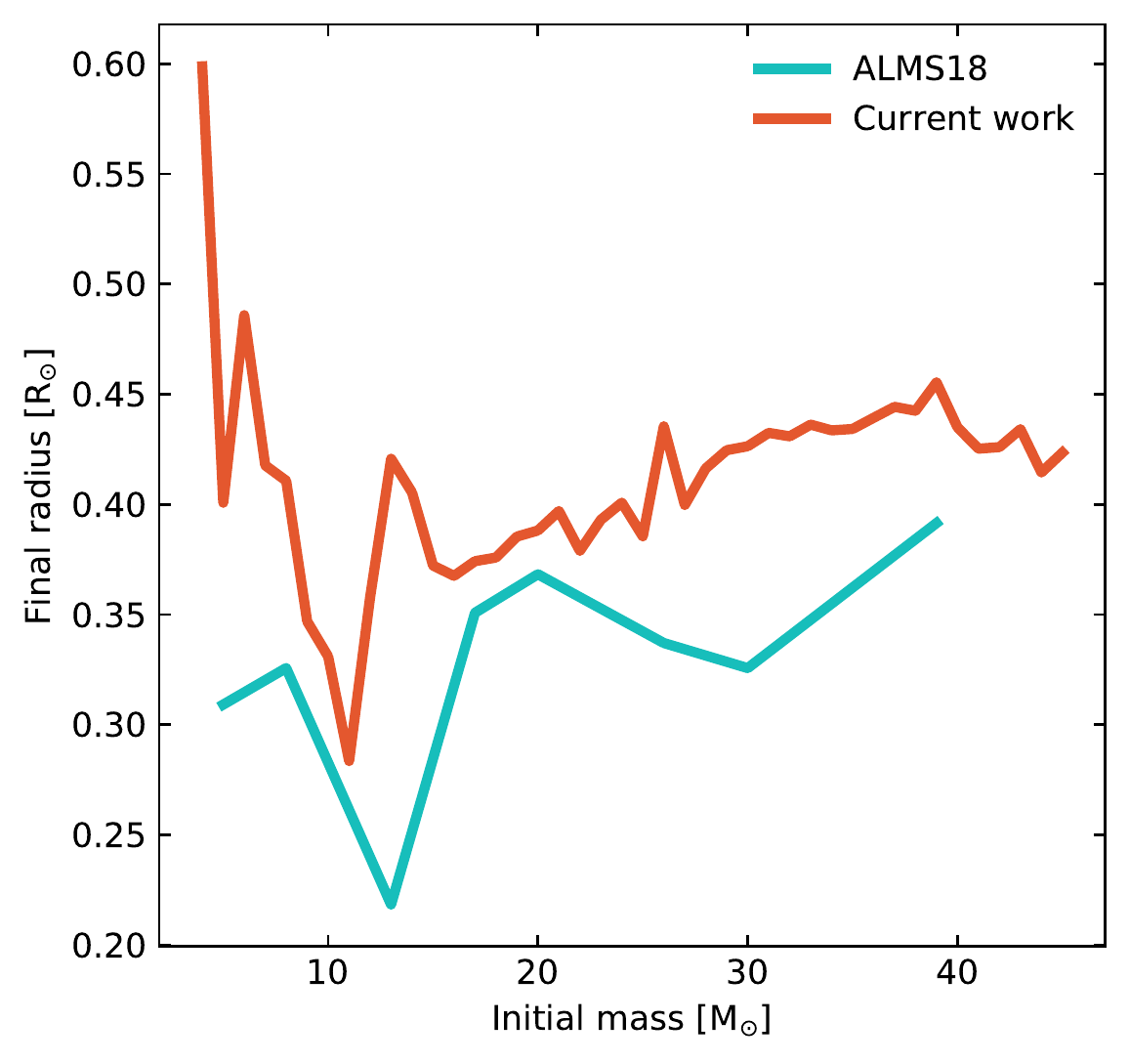}
\caption{Panels show a comparison of different key properties of core collapse models in this study (represented in orange), with the final models from ALMS18 (represented in blue). Panels show the final masses (top), final helium masses in the envelope (excluding the helium formed in the core due to photodisintegration of iron, middle), and final radii (bottom), as a function of initial mass.}\label{fig:comp}
\end{figure}

This is reflected in the top panel of Figure \ref{fig:comp}, where we compare the final masses of previous models to those presented here. Note that, although the current models are precisely at core collapse, whereas the previous ones are calculated at a time shortly before, the remaining lifetime before of previous models is of the order of a few seconds, such that the final masses are well defined. Differences in final masses will become important for the explosive event that follows from this evolutionary channel, since they will determine the ejecta mass, the boundary between core collapse and the onset of pulsational pair instability, as well as the amount of radiation produced by CSM interaction. This is further discussed in Section \ref{sec:obs}.

The middle panel of Figure \ref{fig:comp} reflects the difference in final total helium mass in the envelope. The helium content in the envelope is higher in most cases in the current work due to the higher final masses retained by the models at core collapse. Despite the considerable difference in final mass, reaching up to 25\% in some cases, particularly at higher initial masses, all of these models retain less than 0.013 $\mso$ of helium in their envelopes. We do not expect helium lines to form in these progenitors since they retain very little helium, and all of the remaining helium is located in a carbon/oxygen rich layer which will lower the probability of exciting helium lines, as opposed to the case of a Ib progenitor that has a pure helium and nitrogen layer \citep{2017A&A...603A..51D,2015MNRAS.453.2189D}. Furthermore, excitation of helium lines depends on the mixing of $^{56}$Ni, which may be differently distributed in the ejecta in case of a successful explosion coming from one of our progenitors, likely resulting in a type Ic SN in any case \citep{2019ApJ...872..174Y}.

As can be seen in the bottom panel of Figure \ref{fig:comp}, final radii are not significantly affected by the additional mass, except for the cases below about 8 $\mso$. This is not unexpected since the radius is determined by a competition of the the increasingly stronger neutrino losses stars experience during the late evolutionary stages, and the formation of a helium burning shell. In lower mass models, the relatively larger helium content in the envelope ignites earlier than in the ALMS18 models, and due to the larger lifetime, the helium burning shell manages not only to halt the neutrino-driven contraction, but to produce an expansion of the envelope.

\begin{figure}[h!]
\epsscale{0.95}
\plotone{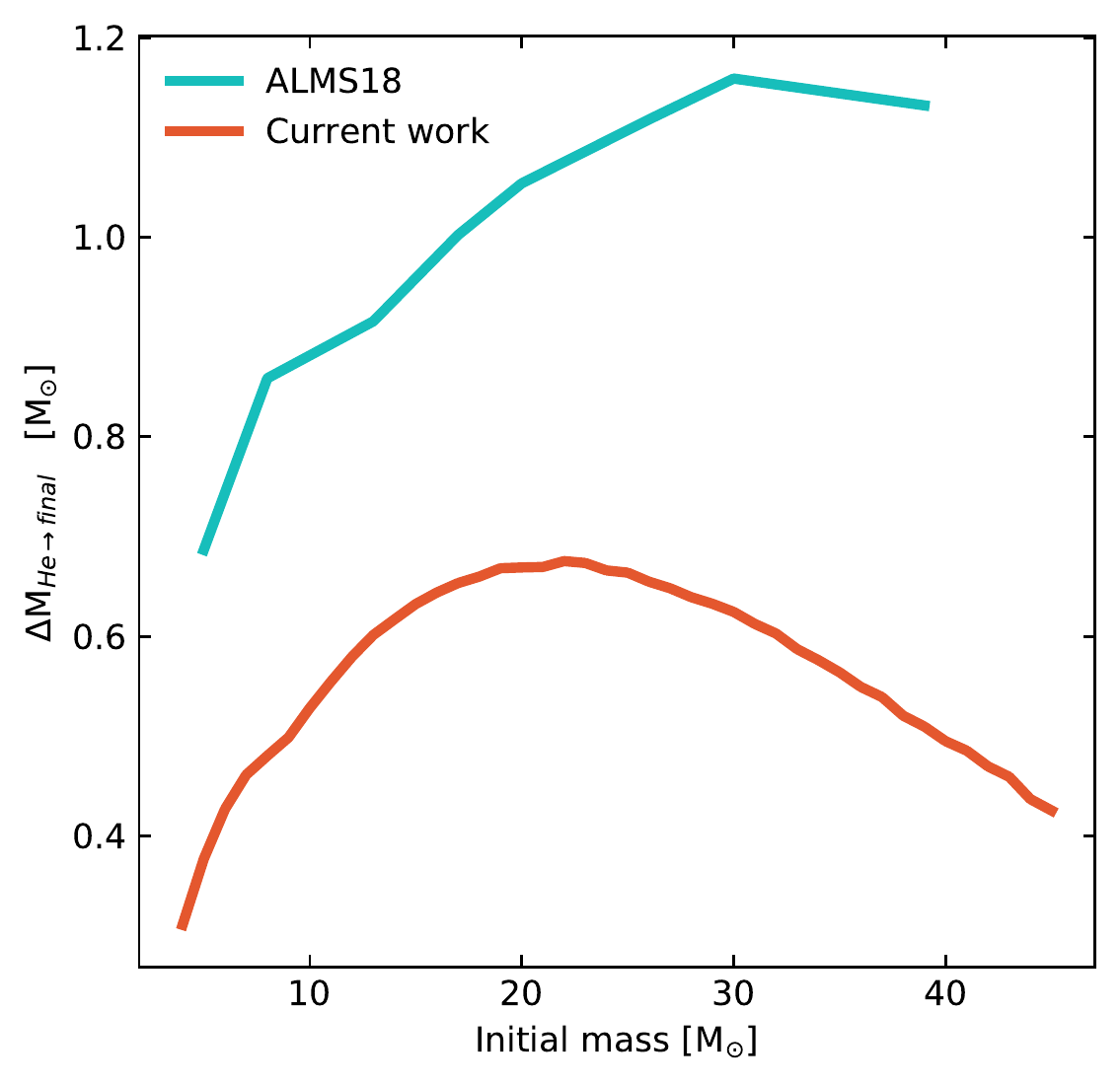}
\caption{Mass lost between core helium depletion and core collapse ($\Delta \text{M}_{\text{He} \rightarrow \text{final}}$), representative of the CSM mass immediately close to the progenitors at core collapse, as a function of initial mass. Comparison between core collapse models from this study (orange) and the final models from ALMS18 (blue).}\label{fig:deltam}
\end{figure}

Another important consequence of a more efficient angular momentum loss is that the immediate CSM mass --produced in the final $\approx$ 1000 years by the combined effect of neutrino-driven contraction and fast rotation-- will be smaller, and the ejecta more massive. The CSM mass, quantified by $\Delta \text{M}_{\text{He} \rightarrow \text{final}}$, the mass lost from core helium depletion until core collapse, as discussed in detail by ALMS18, has to remain close to the star since it is lost by centrifugal acceleration, which decays quickly as the mass moves away from the star, and cannot be accelerated by radiation. A comparison between these quantities is shown in Figure \ref{fig:deltam}. It is striking that some of these CSM masses are reduced by about half by the different treatment of angular momentum losses, but they will still play a significant role in some of the expected transients (see Section \ref{sec:obs}).

\section{Results}\label{sec:results}

As detailed in Section \ref{sub:old}, the global properties and conclusions from ALMS18 remain valid after re-performing the simulations with the changes described in Section \ref{sub:new}. The main parameters of our pre-SN models are summarized in Table \ref{tab:table1}. This Section is divided as follows: In Section \ref{sub:preSN} we discuss the general evolution from ZAMS to core collapse of our models, and in Section \ref{sub:precollapse} we discuss the pre-collapse core properties and explodability of fast rotating stars.

\subsection{Pre-collapse evolution}\label{sub:preSN}

Similar to the simulations in ALMS18, we find that the evolution of fast rotating stars at low metallicity is well described by previous works of quasi-chemically homogeneous stars \citep[e.g.][]{2011A&A...530A.115B} during the main sequence, but late evolution differs significantly from previous studies in a number of ways. The most notorious difference is that rotational mixing and mass loss work together to leave no hydrogen on the surface at the end of the main sequence, and only a small amount of helium in the envelope at the end of core helium burning.

This causes the helium shell that is left to ignite later and be weaker, thus preventing expansion of the stellar radius due to the so-called \textit{mirror principle} \citep{1990sse..book.....K}, and leading to a continuous contraction from ZAMS to core collapse. This contraction proceeds in a thermal time scale, and it becomes accelerated from the end of helium burning by neutrino cooling in the core.

Contraction in these models leads to an increase in surface rotational velocity, which is limited by the critical value, which in turn leads to intense mass loss. In particular during the last few thousand years these stars experience the loss of about 0.5 $\mso$ (see Figure \ref{fig:deltam}, Table \ref{tab:table1}), which cannot be accelerated far from the star and will thus likely interact with the SLSN/lGRB ejecta.

\begin{figure}[h!]
\plotone{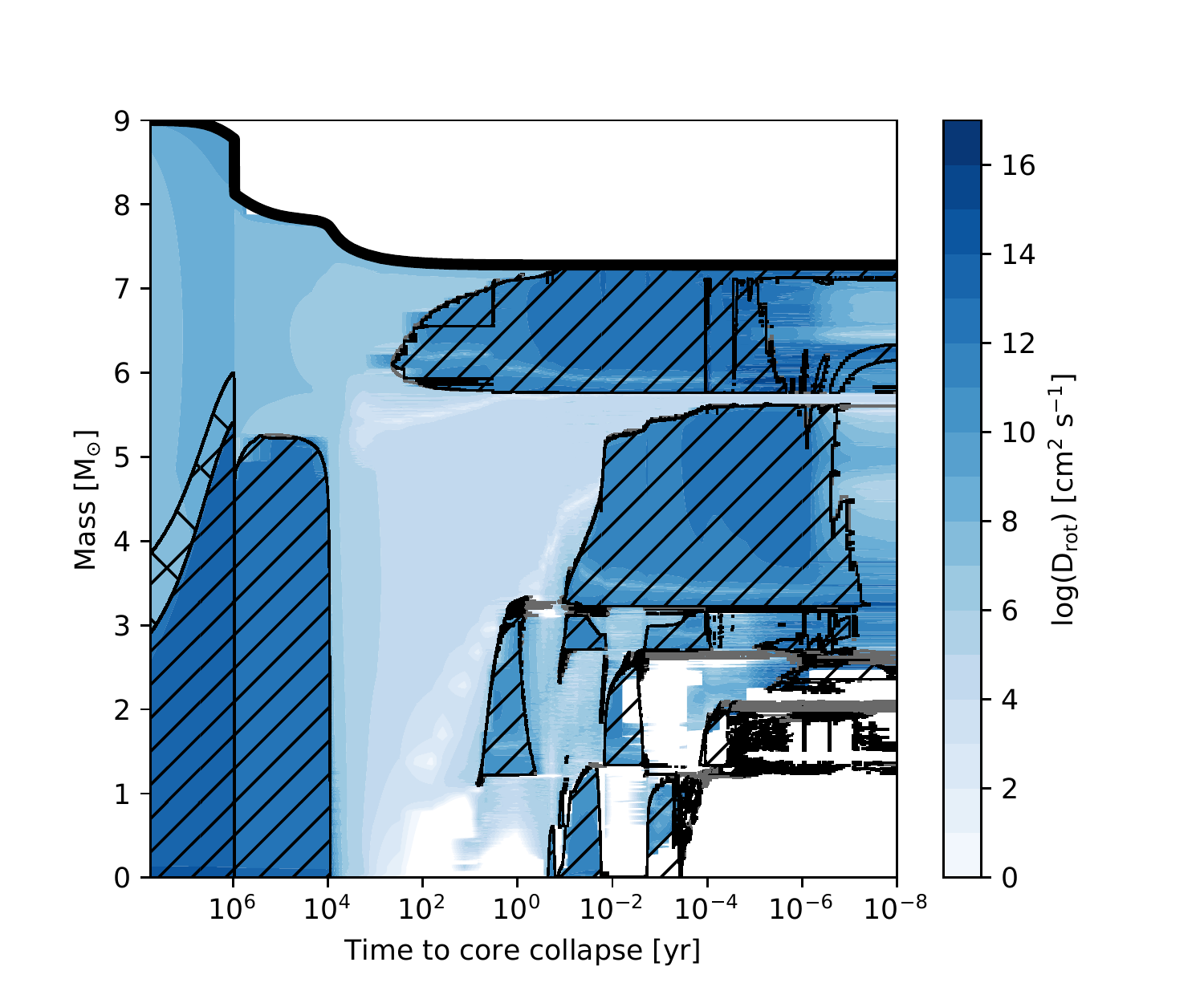}
\caption{Kippenhahn diagram following the efficiency of rotational mixing, and the structure of convective and overshooting regions of the 9 $\mso$ evolutionary calculation from ZAMS to core collapse, as a function of the time remaining before core collapse. Color denotes the diffusion coefficient due to rotational mixing (dominated by Eddington-Sweet circulation). Regions hatched with diagonals denote convective regions, whereas the region hatched with perpendicular lines, above the hydrogen burning core, denotes an overshooting region.}
\label{fig:mix}
\end{figure}

Due to the strong rotational mixing, helium core masses and carbon-oxygen core masses are determined almost exclusively by the mass of the entire star at the end of hydrogen and helium burning, respectively, since no strong composition gradient is present at these times. Only after, during core carbon burning, a distinct core-envelope structure is formed, inhibiting mixing through regions with newly formed composition gradients. This, together with the remaining mass at carbon ignition, determines the core structure in the remainder of the stellar life. This is illustrated in Figure \ref{fig:mix}, where the evolution of the diffusion coefficient for rotational mixing of our 9 $\mso$ model (sum of all effects included in the calculations, but dominated by Eddington-Sweet circulation) is compared to the locations of the convective and overshooting regions during the evolution. It illustrates how the strength of rotational mixing is reduced by the composition gradient established by carbon burning, but it still has a role in shaping the chemical gradient and size of the subsequent carbon, neon and oxygen burning shells in and close to the core by mixing the otherwise static radiative regions between the convective shells.

\begin{deluxetable*}{cccccccccccc}
\tablecaption{Initial mass, final mass, mass lost between ZAMS and the onset of core helium burning, mass lost between the end of helium burning and core collapse of evolutionary sequences in this work, and key parameters of their respective core collapse models. The compactness parameter $\xi_{2.5}$, and the values of M$_4$ and $\mu_4$ as defined by Equations \ref{eq:xi}, \ref{eq:m4} and \ref{eq:mu4} respectively. \label{tab:table1}}
\tablecolumns{7}
\tablewidth{0pt}
\tablehead{
\colhead{$\text{M}_{\text{init}}$}   & \colhead{$\text{M}_{\text{fin}}$ }   & \colhead{$\Delta \text{M}_{\text{H} \rightarrow \text{He}}$}\tablenotemark{a}& \colhead{$\Delta \text{M}_{\text{He} \rightarrow \text{final}}$}\tablenotemark{b} & \colhead{He mass}\tablenotemark{c}& \colhead{Y$_{\text{surf}}$}& \colhead{$\bar{j}_{1.5 \mso}$} & \colhead{$\bar{j}_{2 \mso}$} & \colhead{$\bar{j}_{5 \mso}$} & \colhead{$\xi_{2.5}$ } & \colhead{M$_4$} &  \colhead{$\mu_4$}
\\
\colhead{[$\mso$]} & \colhead{[$\mso$]} & \colhead{[$\mso$]} & \colhead{[$\mso$]}                      & \colhead{[$\mso$]}  & \ & \colhead{[10$^{15}$ \text{cm}$^2$ \text{s}$^{-1}$]} & \colhead{[10$^{15}$ \text{cm}$^2$ \text{s}$^{-1}$]} & \colhead{[10$^{15}$ \text{cm}$^2$ \text{s}$^{-1}$]} & \ & \ &       
}
\setlength\extrarowheight{-1.6pt}
\startdata
4            & 3.31         & 0.20  & 0.32      & 0.013          & 0.10             & 0.51  & 0.91  & 9.07  & 0.13        & 1.77  & 0.063   \\
5            & 4.10         & 0.22  & 0.40      & 0.010          & 0.09             & 0.50  & 1.65  & 20.9  & 0.20        & 1.54  & 0.136   \\
6            & 4.89         & 0.25  & 0.46      & 0.011          & 0.09             & 1.22  & 2.07  & 38.1  & 0.19        & 1.82  & 0.075   \\
7            & 5.68         & 0.30  & 0.49      & 0.012          & 0.08             & 1.07  & 2.13  & 25.1  & 0.13        & 1.57  & 0.062   \\
8            & 6.48         & 0.36  & 0.52      & 0.012          & 0.08             & 1.11  & 2.20  & 20.6  & 0.14        & 1.65  & 0.058   \\
9            & 7.28         & 0.42  & 0.54      & 0.003          & 0.08             & 1.22  & 2.30  & 11.5  & 0.68        & 2.36  & 0.183   \\
10           & 8.06         & 0.49  & 0.56      & 0.004          & 0.08             & 3.01  & 4.05  & 20.0  & 0.50        & 2.24  & 0.152   \\
11           & 8.84         & 0.56  & 0.58      & 0.006          & 0.08             & 2.42  & 3.39  & 15.6  & 0.42        & 2.12  & 0.123   \\
12           & 9.61         & 0.63  & 0.60      & 0.009          & 0.08             & 2.20  & 3.31  & 14.1  & 0.23        & 1.88  & 0.068   \\
13           & 10.37        & 0.71  & 0.62      & 0.011          & 0.08             & 1.24  & 2.76  & 12.6  & 0.21        & 1.86  & 0.073   \\
14           & 11.13        & 0.78  & 0.63      & 0.010          & 0.08             & 1.12  & 2.46  & 11.0  & 0.31        & 2.07  & 0.092   \\
15           & 11.89        & 0.87  & 0.65      & 0.011          & 0.08             & 1.36  & 2.76  & 11.1  & 0.37        & 2.15  & 0.107   \\
16           & 12.64        & 0.95  & 0.66      & 0.010          & 0.08             & 1.46  & 2.86  & 9.73  & 0.54        & 2.34  & 0.197   \\
17           & 13.39        & 1.04  & 0.67      & 0.010          & 0.08             & 1.33  & 2.58  & 8.96  & 0.59        & 2.45  & 0.163   \\
18           & 14.15        & 1.12  & 0.68      & 0.010          & 0.08             & 1.34  & 2.64  & 8.62  & 0.62        & 2.49  & 0.169   \\
19           & 14.89        & 1.21  & 0.69      & 0.010          & 0.08             & 1.19  & 2.29  & 8.22  & 0.65        & 2.27  & 0.262   \\
20           & 15.64        & 1.29  & 0.69      & 0.009          & 0.09             & 0.98  & 1.95  & 7.93  & 0.66        & 2.28  & 0.278   \\
21           & 16.39        & 1.38  & 0.69      & 0.008          & 0.09             & 1.07  & 2.17  & 7.95  & 0.66        & 2.15  & 0.304   \\
22           & 17.12        & 1.47  & 0.70      & 0.007          & 0.09             & 1.06  & 2.02  & 8.24  & 0.63        & 2.01  & 0.277   \\
23           & 17.86        & 1.56  & 0.70      & 0.008          & 0.09             & 1.26  & 2.46  & 8.88  & 0.54        & 2.04  & 0.285   \\
24           & 18.60        & 1.65  & 0.69      & 0.008          & 0.09             & 0.43  & 0.93  & 8.66  & 0.53        & 1.71  & 0.259   \\
25           & 19.33        & 1.76  & 0.69      & 0.007          & 0.09             & 0.33  & 0.68  & 8.88  & 0.47        & 1.75  & 0.186   \\
26           & 20.05        & 1.87  & 0.68      & 0.010          & 0.09             & 1.75  & 3.19  & 9.29  & 0.36        & 1.93  & 0.115   \\
27           & 20.78        & 1.98  & 0.67      & 0.007          & 0.09             & 1.30  & 2.46  & 9.52  & 0.43        & 1.85  & 0.153   \\
28           & 21.50        & 2.10  & 0.66      & 0.007          & 0.09             & 1.56  & 2.95  & 9.70  & 0.40        & 1.92  & 0.133   \\
29           & 22.22        & 2.24  & 0.66      & 0.008          & 0.09             & 1.77  & 3.43  & 10.4  & 0.43        & 2.07  & 0.124   \\
30           & 22.95        & 2.36  & 0.65      & 0.008          & 0.09             & 1.34  & 2.63  & 9.54  & 0.43        & 2.08  & 0.129   \\
31           & 23.66        & 2.48  & 0.64      & 0.008          & 0.09             & 1.17  & 2.46  & 9.51  & 0.43        & 2.04  & 0.134   \\
32           & 24.38        & 2.62  & 0.63      & 0.008          & 0.09             & 2.01  & 3.73  & 10.5  & 0.52        & 2.26  & 0.137   \\
33           & 25.10        & 2.75  & 0.61      & 0.008          & 0.09             & 2.07  & 3.75  & 9.97  & 0.55        & 2.28  & 0.138   \\
34           & 25.82        & 2.90  & 0.60      & 0.007          & 0.10             & 1.97  & 3.55  & 10.0  & 0.60        & 2.35  & 0.153   \\
35           & 26.53        & 3.04  & 0.59      & 0.008          & 0.10             & 1.57  & 2.95  & 9.44  & 0.57        & 2.33  & 0.149   \\
36           & 27.24        & 3.18  & 0.58      & 0.007          & 0.10             & 1.55  & 2.82  & 8.84  & 0.62        & 2.37  & 0.154   \\
37           & 27.95        & 3.33  & 0.57      & 0.007          & 0.10             & 1.86  & 3.23  & 9.35  & 0.68        & 2.43  & 0.159   \\
38           & 28.67        & 3.49  & 0.55      & 0.007          & 0.10             & 1.65  & 2.88  & 9.01  & 0.75        & 2.50  & 0.178   \\
39           & 29.37        & 3.64  & 0.54      & 0.008          & 0.10             & 2.28  & 3.78  & 9.44  & 0.75        & 2.54  & 0.183   \\
40 & 30.08 & 3.8  & 0.53 & 0.008 & 0.10 & 2.08 & 3.32 & 9.92 & 0.78 & 2.56 & 0.184 \\
41 & 30.78 & 3.97 & 0.52 & 0.007 & 0.10 & 1.49 & 2.47 & 8.33 & 0.77 & 2.6  & 0.173 \\
42 & 31.49 & 4.14 & 0.5  & 0.007 & 0.10 & 2.24 & 3.50 & 8.81 & 0.83 & 2.63 & 0.209 \\
43 & 32.18 & 4.3  & 0.49 & 0.007 & 0.10 & 2.06 & 3.25 & 9.22 & 0.84 & 2.71 & 0.219 \\
44 & 32.89 & 4.49 & 0.47 & 0.007 & 0.10 & 2.31 & 3.56 & 9.05 & 0.85 & 2.78 & 0.238 \\
45 & 33.59 & 4.67 & 0.46 & 0.006 & 0.10 & 1.95 & 3.01 & 8.54 & 0.85 & 2.84 & 0.215 
\enddata
\tablenotetext{a}{Mass lost from between H and He burning, defined as $\Delta \rm{M}_{\rm H \rightarrow He} = \int_{T8_{\rm c} > 0.5}^{T8_{\rm c} < 1.2} \dot M \, dt$}
\tablenotetext{b}{Mass lost from He core depletion to the end of the simulation}
\tablenotetext{c}{In the envelope}
\end{deluxetable*}

\subsection{Properties at core collapse}\label{sub:precollapse}

With an increased resolution in mass we are able to set a limit to the onset of pulsational pair instability between initial masses of 45 and 46 $\mso$, and to determine with higher accuracy the fate of these evolutionary sequences at and after core collapse. This, however, depends on the mass of the star at the onset of carbon burning \citep{2017ApJ...836..244W}, and on the amount of carbon at the end of helium burning \citep{2019ApJ...887...53F}, and therefore is uncertain inasmuch as the loss of angular momentum, and reaction rates remain uncertain.

The amount of angular momentum that is retained in the interior of the modelled stars leads to the expectation that they could form a collapsar, resulting in a lGRB, in the case that core collapse leads to BH formation. Otherwise, if core collapse leads to the formation of a NS, its spin and magnetic field will be consistent with those expected from SLSNe.

To assess which parts of the parameter space correspond to each of these possible fates, we make the ansatz that those models that we predict successfully explode as neutrino-driven SNe, would produce a SLSN, since this NS would be very fast rotating and highly magnetized, comparable to the magnetars required to produce SLSNe. On the other hand, we assume that explosions that we predict would not successfully explode, but instead form BHs, would produce collapsars due to the high content of angular momentum in our models. This would in turn power both a lGRB and its accompanying type Ic-BL SN. To determine which core-collapse models correspond to either SLSNe or lGRBs, we make use of results obtained for neutrino-driven core collapse of non-rotating SN progenitors.
Applying explosion criteria for neutrino-driven SNe is evidently an oversimplification, but
can nonetheless be justified as a first estimate.
One concern is that the mechanism for shock
revival in SLSNe may rely on magnetohydrodynamic
effects to begin with, so that success and failure
are determined by different criteria.
However, this is by no means certain. Recent SN simulations suggest that neutrino-driven
explosions can occur quite readily in
rapidly rotating massive progenitors
\citep{2016MNRAS.461L.112T,2018ApJ...852...28S,2020MNRAS.492.4613O} thanks to support by centrifugal
forces and possibly a corotation instability.
Even if part or most of the explosion energy
is ultimately delivered by magnetohydrodynamic
effects in SLSNe, there may well be a neutrino-driven 
``precursor'' explosion, whose occurrence will be 
governed by similar explodability criteria as in the
non-rotating case, though the criteria will be relaxed 
\citep{2018ApJ...852...28S} due to rotational support.
Moreover, regardless of the nature of the
SN mechanism, explosions will be inhibited
or cut off by similar factors, e.g., high
pre-shock ram pressure, a high binding energy
of the shells surrounding the core, and an unduly
high core mass that leaves little time for a
successful explosion to develop before ongoing
accretion onto the NS results in
collapse to a BH. Hence we expect that 
common explodability criteria for non-rotating
progenitors will remain qualitatively useful
even for magnetorotational SNe.

The importance of the core compactness in SN models --a measure of the gravitational binding energy near the core of pre-SN stars-- in determining the final fate of a stellar model, has been pointed out as a possible tool to determine whether they will successfully explode \citep{2011ApJ...730...70O,2012ApJ...757...69U,2014ApJ...783...10S,2016MNRAS.460..742M,2018ApJ...860...93S}.

\begin{figure}[h!]
\plotone{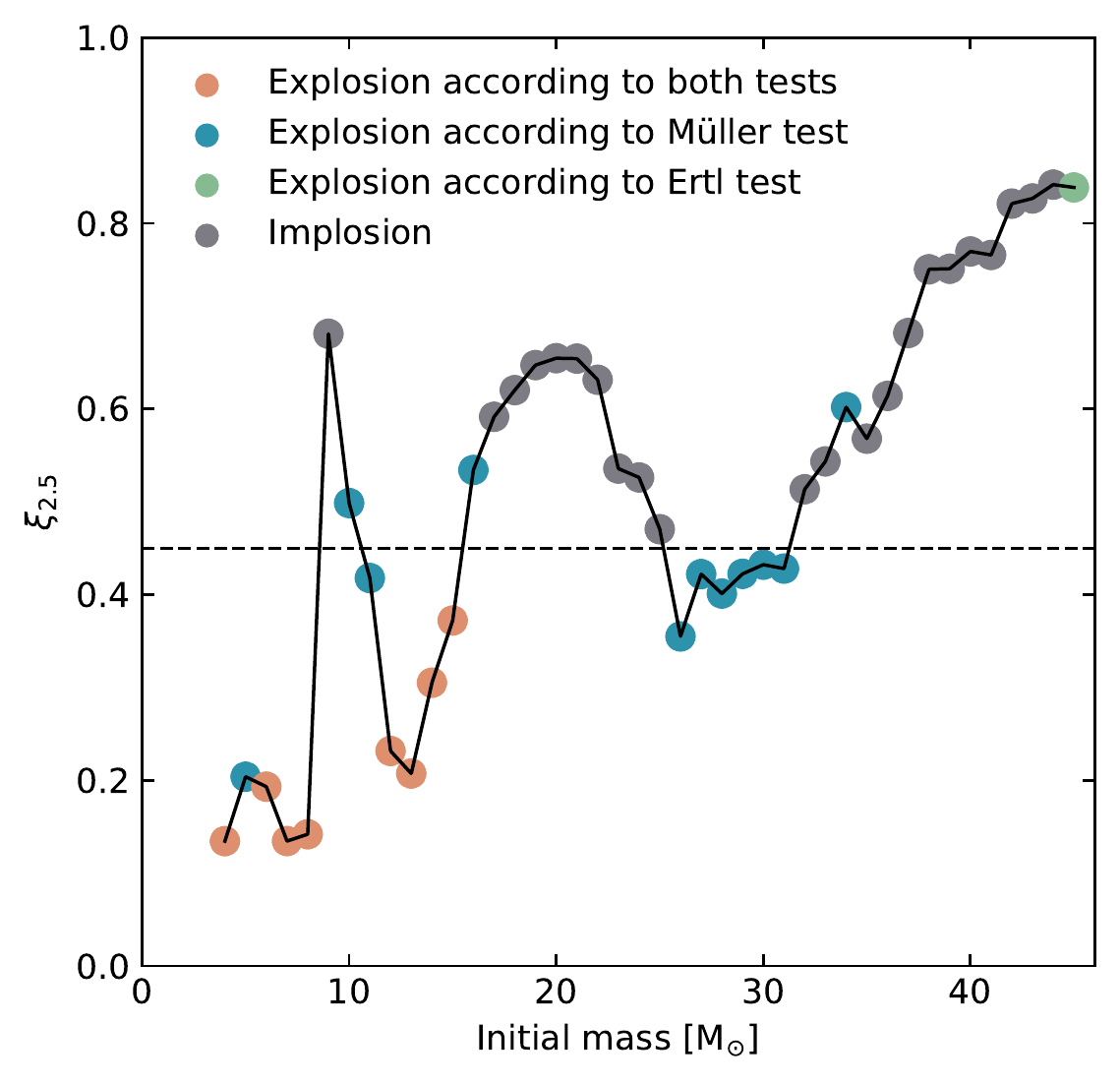}
\caption{Compactness parameter measured at 2.5 $\mso$ of the core collapse models in this study, as a function of their initial mass. The dotted line at $\xi_{2.5}=0.45$ separates models that might explode (below the line) or implode (above the line) according to \cite{2011ApJ...730...70O}, orange points indicate models that are predicted to explode according to the explodability the \cite{2016MNRAS.460..742M} and \cite{2016ApJ...818..124E} tests, blue points are models that are predicted to explode according the \cite{2016MNRAS.460..742M} test, but not by the \cite{2016ApJ...818..124E} test, the green point is a model that is predicted to explode according to the \cite{2016ApJ...818..124E} test but not the \cite{2016MNRAS.460..742M} test, and grey points are models that would not successfully explode according to both tests.}\label{fig:compactness}
\end{figure}

The core compactness was defined by \cite{2011ApJ...730...70O} as
\begin{equation}\label{eq:xi}
\xi_{\text{M}} = \frac{\text{M}/\mso}{\text{R}(\text{M}_{\text{bary}}=\text{M})/1000 \ \text{km}},
\end{equation}
and it was recognized as an indicator of whether the collapse of a non-rotating stellar core leads to a successful explosion (where neutrino winds are the cause of the explosion), or conversely ends up with the formation of a BH. It was found by \cite{2014ApJ...783...10S} that $\xi_{\text{M}}$ is well determined when measured at a mass coordinate of 2.5 $\mso$ at core collapse, which is defined as the point where the infall velocity in the core reaches 1000 $\kms$. . The fate of a core collapse event is then approximately determined by whether $\xi_{2.5}$ is smaller than 4.5, which leads to a successful neutrino-driven explosion; or larger, which leads to the formation of a BH . The compactness parameter of our models at core collapse is illustrated in Figure \ref{fig:compactness}. Although this test is not sufficient to accurately predict whether a stellar model will explode or not \citep{2016ApJ...818..124E,2016MNRAS.460..742M}, more so for fast rotators that have additional sources of energy that could help a successful explosion, it still provides useful information of the structure of a stellar core in the pre-SN stage.

To improve on the predictions of Equation \ref{eq:xi}, \cite{2016ApJ...818..124E} propose a  method to determine the fate of a core-collapse event with a two parameter test. They define the parameters
\begin{equation}\label{eq:m4}
\text{M}_4 = \text{m}(\text{s}=4)/\mso,
\end{equation}
where $m$ is the Lagrangian mass coordinate and s is the specific entropy in units of $k_B$, and
\begin{equation}\label{eq:mu4}
\mu_4 =\left. \frac{ \text{dm} / \mso}{\text{dr}/ 1000 \ \text{km}} \right|_{\text{s}=4},
\end{equation}
the mass gradient evaluated at $M_4$, evaluated in practice by setting $dm = 0.3 \mso$ and dividing by the change in radius between $M_4$ and $M_4 + dm$. \cite{2016ApJ...818..124E} provide a calibration of the two parameters where they found the boundary between successful and failed explosions. We evaluate these parameters in our models with the default parameters in \cite{2016ApJ...818..124E}, showing the exploding models according to this test as orange dots in Figure \ref{fig:compactness}.

A more sophisticated method to determine explodability was provided by \cite{2016MNRAS.460..742M}, who created a semi-analytic model of the formation of a proto-NS and how it grows by accreting material from its surroundings, and injects a fraction of its neutrino luminosity into the outflowing layers above it, parametrizing the onset of the explosion with mass, radius, density and entropy distribution as inputs from the stellar model at core-collapse, and yielding not only the explodability of a certain model, but also providing other parameters of the explosion. 

Since we expect that rotation and magnetic
fields increase the likelihood for a
successful explosion, we adopted slightly more
optimistic values for some of the model parameters.
In particular, we employed a smaller value for the efficiency factor for conversion of accretion energy into $\nu$ luminosity, setting $\zeta =$ 0.7 instead of 0.8, and a longer cooling time-scale for 1.5 $\mso$ NS, setting it to $\tau_{1.5} =$ 1.5 s instead of 1.2 s. We found that the dominating effect is to set a shorter $\tau_{1.5}$, which corresponds to injecting energy more quickly into the layers above the collapsing core. These choices correspond to reducing the importance of accretion power in delivering the energy to the gain region, which was done to mimic an energy source that depends on the NS binding energy rather than on the accretion rate, similar to the case we would expect in a magnetorotational mechanism that taps energy from rotation rather than from neutrino emission.

Figure \ref{fig:compactness} shows also the results of this test, pointing out the exploding models with orange and blue dots.
With few exceptions, the semi-analytic model is
compatible with a dividing line at  
$\xi_{2.5}\approx 0.45$ between NS and
BH formation,  which is somewhat higher 
than the threshold of
$\xi_{2.5}\approx 0.3\texttt{-}0.35$ for the
standard case of \citet{2016MNRAS.460..742M},
as intended.

The non-monotonic behavior of the compactness parameter as a function of initial mass, as well as the presence of a few cases of mismatching predictions between different tests, also present in previous studies, can be observed to persist in our fast rotating and efficiently mixed evolutionary sequences. This suggests that if stars undergo quasi-chemically homogeneous evolution and rotational mixing is efficient, they may result in a variety of transient events depending on their core properties, and that the outcome of their explosions is not a monotonic function of the progenitor mass. However, it must be reiterated that these results are approximate and are not calibrated to work with fast rotating stars. Recently, \cite{2020arXiv200210115P} performed a 3D hydrodynamical, neutrino-driven simulation of the core-collapse SN resulting from a 39 $\mso$ model similar to those in ALMS18.  This resulted in a successful SN explosion that produced a massive NS instead of a BH. \citet{2020MNRAS.491.2715B} also
expressed doubts about the anticorrelation of the compactness parameter with explodability based
on a large set of 3D SN simulations.
On the other hand, these simulation results
are also beset with uncertainties and
will not necessarily be closer to reality
than the phenomenological explosion criteria. Especially
explosions for very high compactness
would be difficult to reconcile with the lack of 
high-mass stars  among observationally identified SN progenitors \citep{2015PASA...32...16S}.

\begin{figure}[ht!]
\gridline{\fig{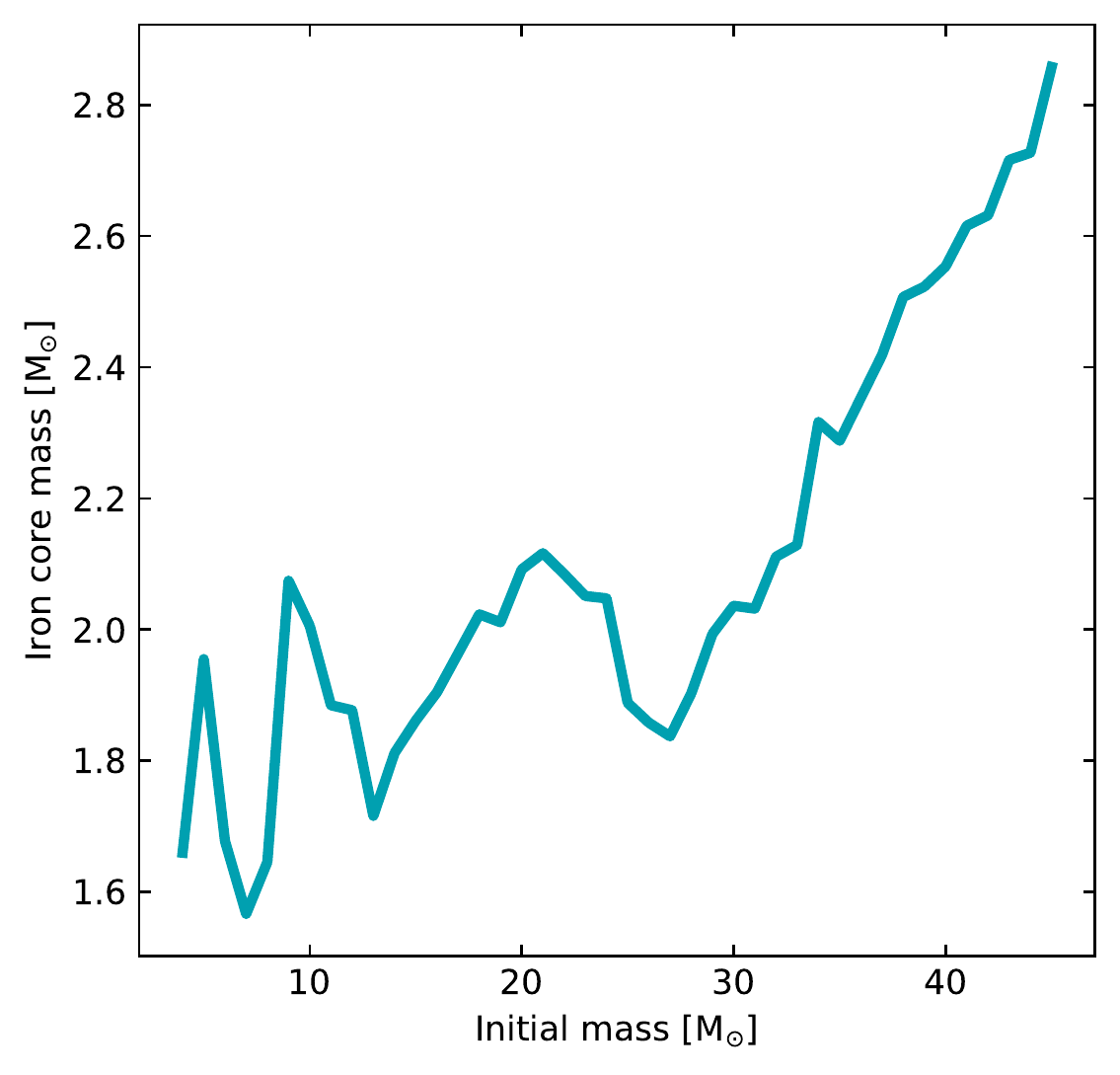}{0.4\textwidth}{}}
\caption{Iron core mass of the core collapse models in this study, as a function of initial mass.}\label{fig:FeCores}
\end{figure}

\begin{figure}[ht!]
\gridline{\fig{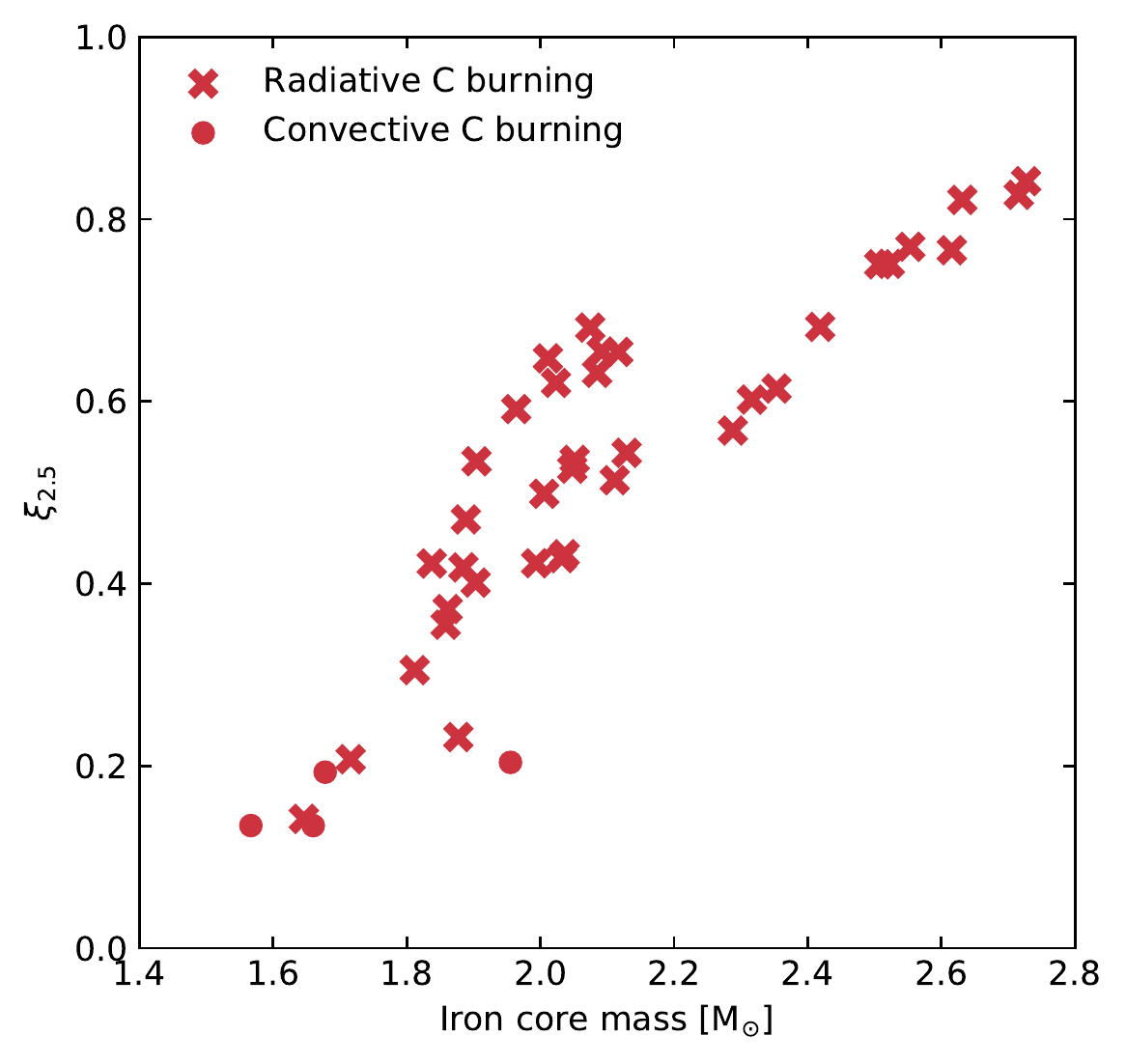}{0.4\textwidth}{}}
\caption{Compactness parameter measured at 2.5 $\mso$ of the core collapse models in this study, as a function of iron core mass at the moment of core collapse. Crosses indicate models where core carbon burning occurred radiatively, whereas points indicate convective core carbon burning.}\label{fig:xi_coremass}
\end{figure}

Compactness (and therefore explodability) seems to correlate closest with iron core mass, which is also a non-monotonic function of initial mass. As seen in Figure \ref{fig:FeCores}, the distribution of iron core masses resembles closely that of $\xi_{2.5}$. This is also reflected in Figure \ref{fig:xi_coremass}, where the correlation between $\xi_{2.5}$ and iron core mass is explicitly demonstrated. Figure \ref{fig:xi_coremass} also shows that models where convective carbon burning occurs tend to have a lower overall compactness, as found by \citealt{2018ApJ...860...93S}.

We can infer from these results that BH formation may be favorable in the mass range from 9 to 11 $\mso$, and in those exceeding around 15 $\mso$, excluding perhaps a window around 25-31 $\mso$, where explosions are not predicted by the \cite{2016ApJ...818..124E} method, but are predicted to explode by the \cite{2016MNRAS.460..742M} method and have relatively low compactness. Furthermore, these models might be affected by centrifugal acceleration, rotationally induced instabilities and the evolution of their magnetic field during collapse, which might result in a different number of explosions in a given mass range.

\begin{figure*}
\gridline{\fig{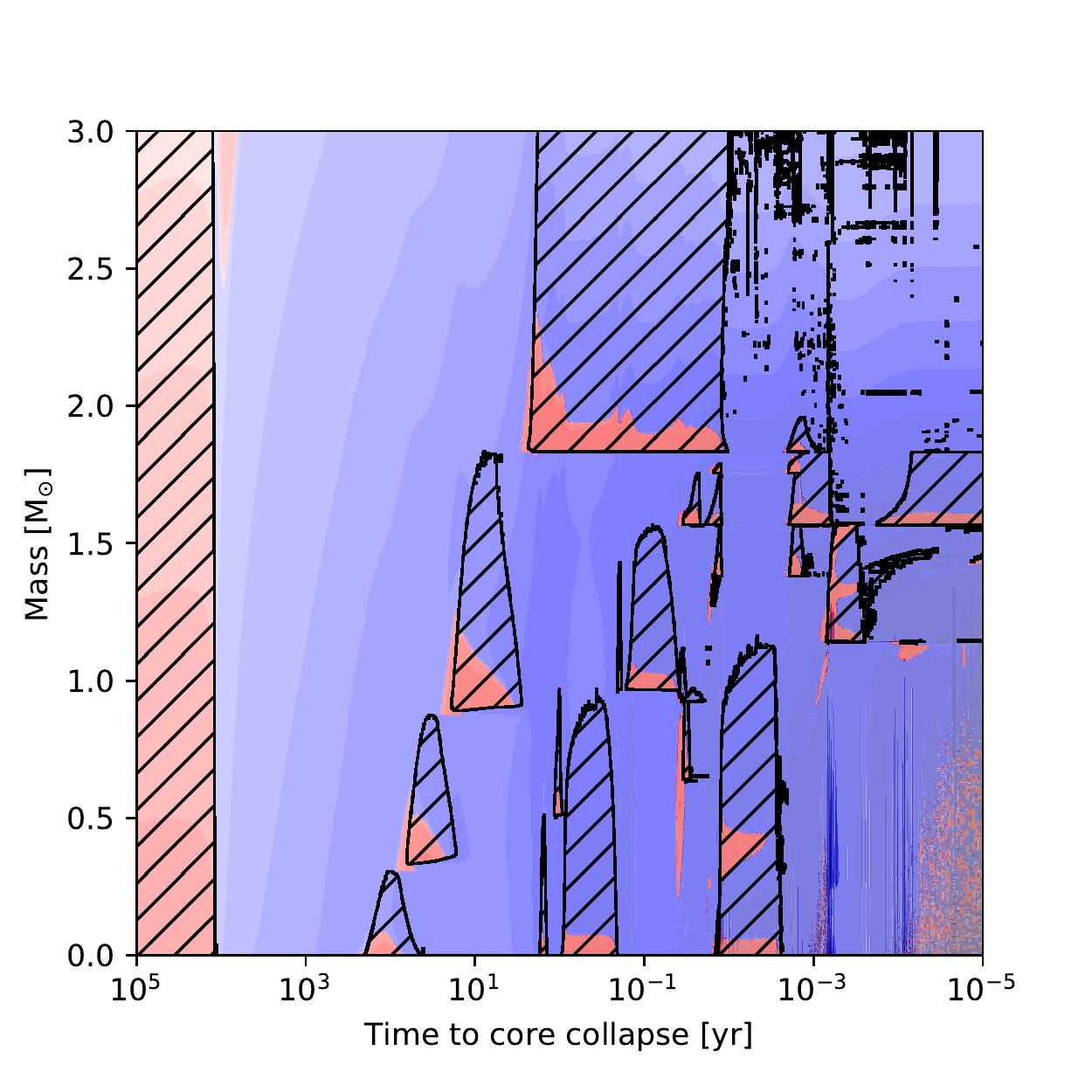}{0.333333\textwidth}{} \fig{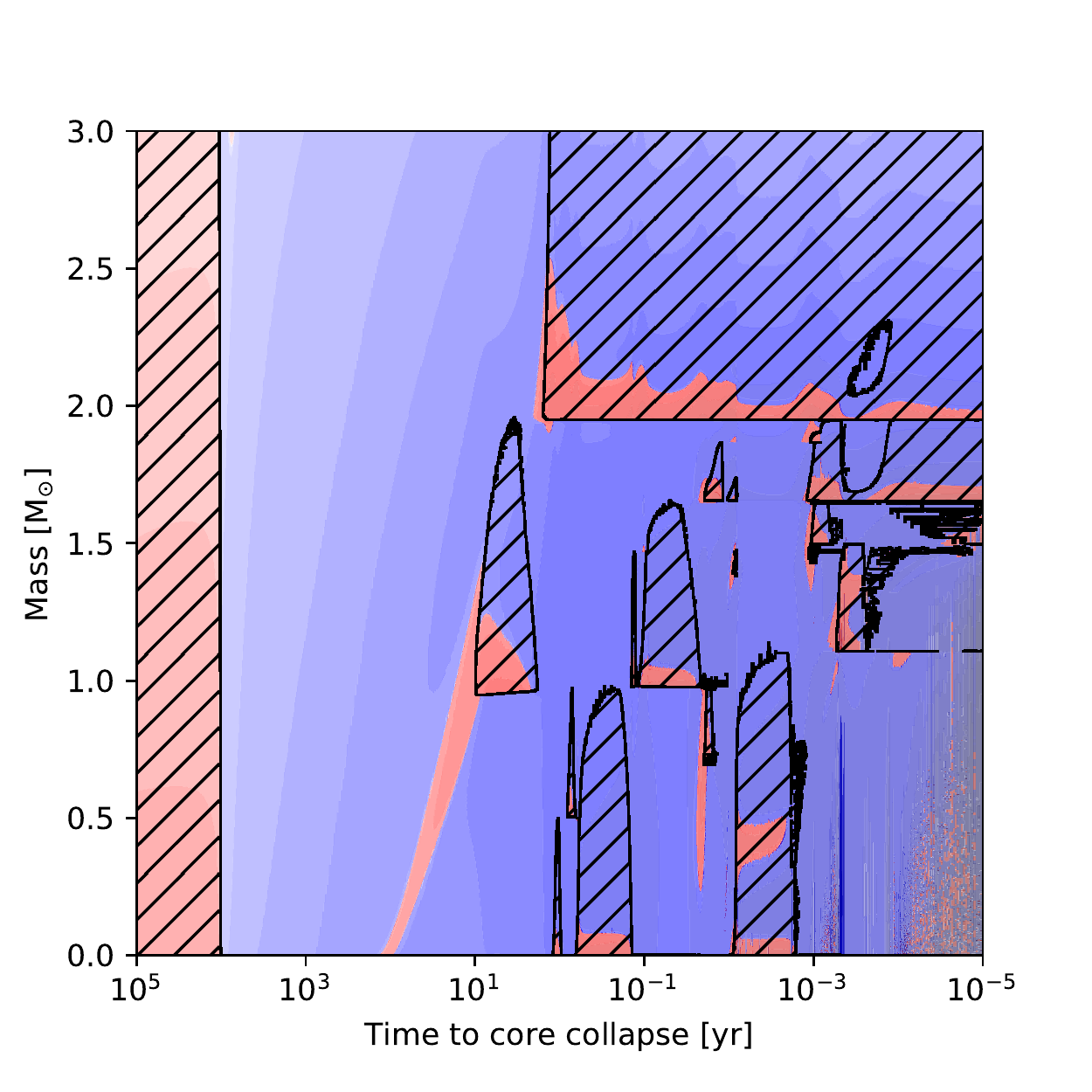}{0.333333\textwidth}{}\fig{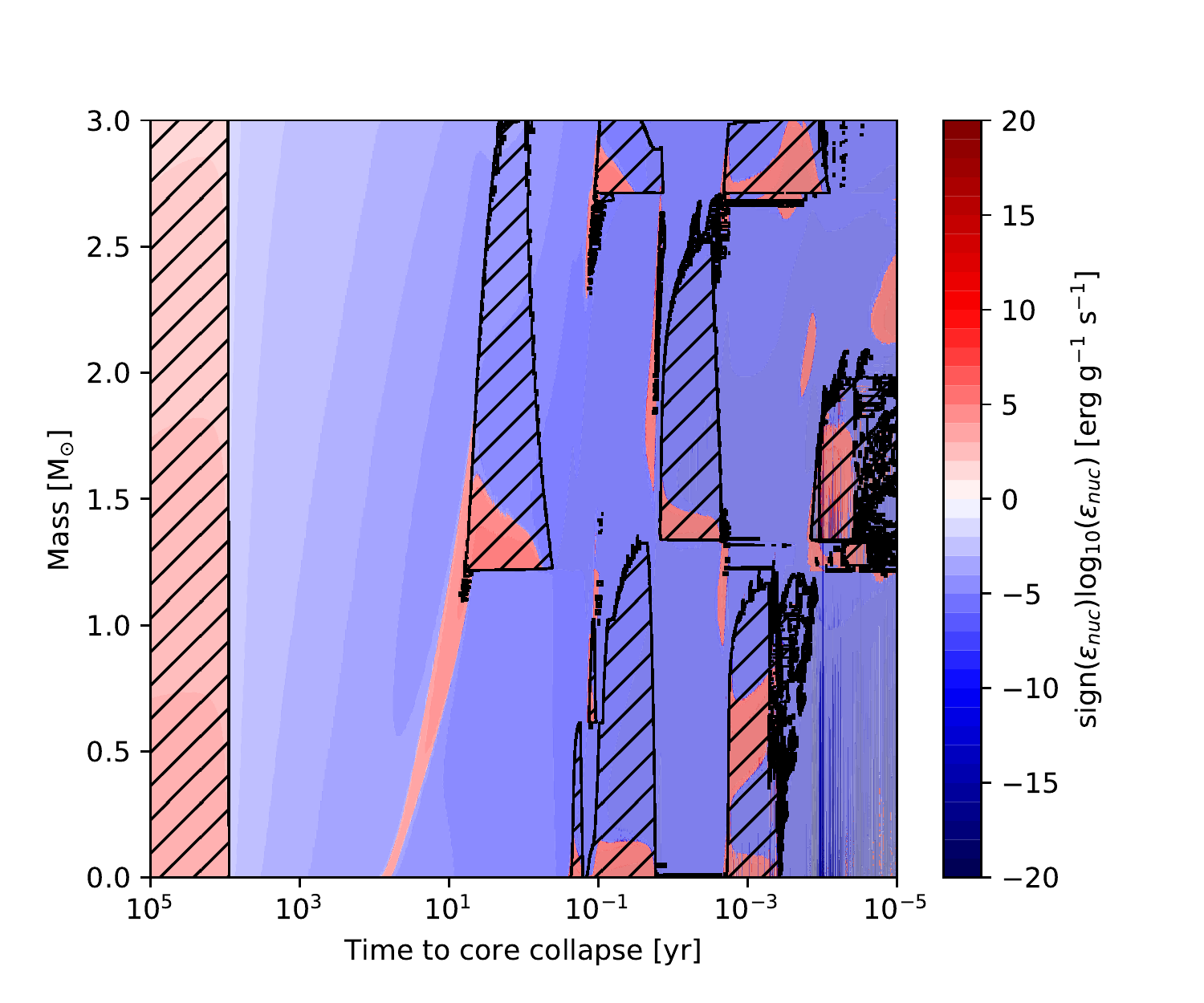}{0.4\textwidth}{}}
\caption{Kippenhahn diagrams following the energy generation/loss rate, and the structure of convective regions as a funciton of time remaining until core collapse, from core helium burning until minutes before core collapse. Represented are the evolutionary sequences with initial masses of 7 (left), 8 (middle) and 9 $\mso$ (right). Color denotes the intensity of energy generation rate (red), and the energy loss rate (blue). Regions hatched with diagonals denote convective regions.}
\label{fig:kips789}
\end{figure*}

This suggests that the parameter space where we expect to form SLSNe powered by a magnetar or lGRBs powered by a collapsar is a non-monotonic function of mass, and windows of explodibility exist in the case of chemically homogeneously evolving stars as well, even if rotational mixing is efficient. However, it is important to reiterate that this threshold for explodibility are calibrated with explosion models powered by neutrino heaeting, and detailed numerical calculations are still required to study the effect of energy deposition by the central engine itself, the evolution of the magnetic fields, and other 3D effects that might become important during the collapse.

Although the non-monotonic behavior of $\xi_{2.5}$ is present in all pre-SN models, some of the differences between the canonical evolutionary channel of single massive stars and chemically homogeneous stars are reflected in the behavior of $\xi_{2.5}$. Despite the low metalicity, rotation and neutrino-driven contraction cause a very strong rotationally enhanced mass loss rate, which ends up accounting for around 0.5 -- 0.7 $\mso$ of material in the last 1000 years of evolution (depending on the loss of angular momentum, see Section \ref{sub:new}), as well as a large fraction of the initial mass during the rest of its evolution, particularly during the contraction after hydrogen and helium burning. Mass loss is already known to have an important effect \citep{2017A&A...603A.118R,2018ApJ...860...93S} in the behavior of $\xi_{2.5}$, since it sets the relation between initial mass and carbon/oxygen core mass, which in turn fixes the boundaries between different behaviors during carbon and oxygen burning.

Models with ZAMS masses between 7 and 14 $\mso$ are expected to have more variations than we currently trace with the resolution in initial mass in our models. As was shown by \cite{2014ApJ...783...10S} and \cite{2018ApJ...860...93S}, solutions to the equations of stellar structure vary significantly with small changes in mass in this regime, due to the transition that occurs between radiative and convective core carbon burning, which is shown in Figure \ref{fig:kips789} to occur in this mass regime, and which was found to have a very strong effect on $\xi_{2.5}$. In our simulations, this transition is found between 7 and 8 $\mso$, but a strong variation of $\xi_{2.5}$ was resolved between 8 and 9 $\mso$ and beyond. The first and second carbon shells disappear between 7 and 8 $\mso$, but between 8 and 9 $\mso$ the third carbon burning shell (following the notation from \cite{2014ApJ...783...10S}) ignites much more intensely, resulting in a convective region that spans almost all of the carbon/oxygen shell, and that burns strongly until the core collapses, as opposed to lower mass models which develop a number of smaller carbon burning shells through their evolution. This, in turn, results in a later ignition of neon/oxygen, allowing the innermost part of the core to contract for a longer time, resulting in a higher value of $\xi_{2.5}$.

Once the core exceeds around 15 $\mso$, however, the core structure becomes more regular: Radiative carbon burning followed by convective neon and oxygen burning shells, a convective region that encompasses the silicon core and the oxygen shell, and a strong carbon burning shell on top that remains on until the end of the evolution, and that is active from the edge of the core to the helium burning shell close to the surface. As each of these regions grows in mass with the total mass of the star, the factors that then determine the compactness are the size of the core, which determines the relative location of the 2.5 $\mso$ mass coordinate where we define the compactness, and the behavior of the silicon burning region and oxygen shells only minutes before core collapse.

\begin{figure}[h!]
\plotone{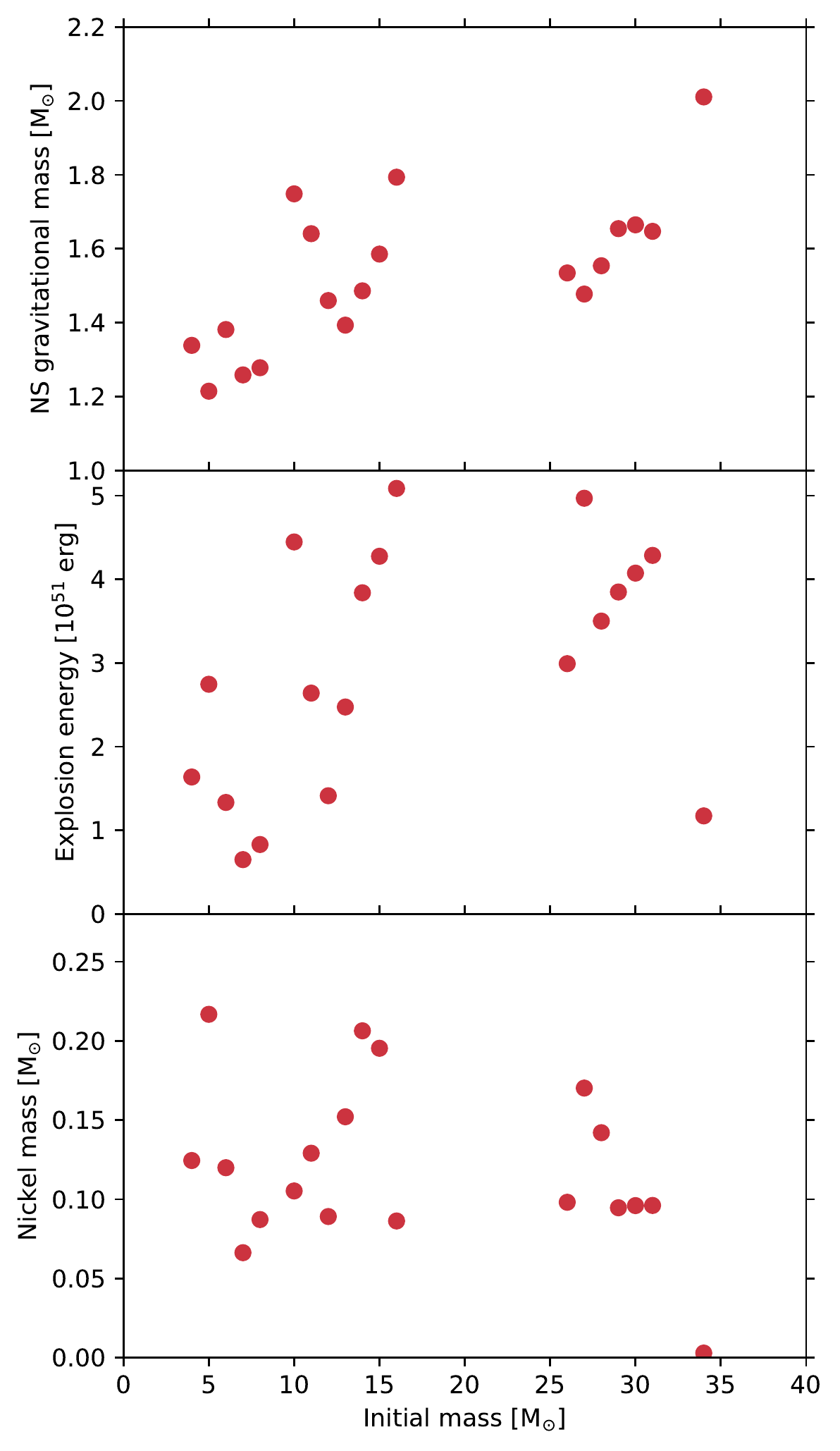}
\caption{NS gravitational mass, explosion energy and $^{56}$Ni mass resulting from SN explosions produced from our core collapse models, as predicted by the \cite{2016MNRAS.460..742M} model, as a function of initial mass.
}\label{fig:panel}
\end{figure}

Using the \cite{2016MNRAS.460..742M} analytical model to determine explodability of these models also yields a number of other interesting parameters. These are summarized in Figure \ref{fig:panel} and in Table \ref{tab:table2}.As expected from the relatively high compactness and high core masses. these models will --on average-- lead to a large amount of energy deposited by neutrinos into the ejecta, as well as producing massive NSs, and large amounts of $^{56}$Ni during the explosion. Strictly speaking,
this only means that we expect
a large energy and ejected nickel mass of a neutrino-driven ``precursor'' to the millisecond magnetar phase, or 
a significant auxiliary contribution of neutrino heating to the explosion energy in a magnetorotational SN.
For this reason, we also list the rotational
energy $E_\mathrm{rot,NS}$ of the NS
in Table~\ref{tab:table2} as a rough estimate
for the energy attainable in a magnetorotational explosion.

\begin{deluxetable*}{cccccccc}
\tablecaption{Explosion parameters inferred from core collapse models in this work analyzed with the \cite{2016MNRAS.460..742M} model, and rotational energy of the resulting NS, total rotational and absolute value of their total energies (sum of gravitational, internal, kinetic and rotational energies) at core collapse.\label{tab:table2}}
\tablecolumns{5}
\tablewidth{0pt}
\tablehead{
\colhead{M$_{\text{init}}$}   &\colhead{E$_{\text{exp}}$\tablenotemark{a}} & \colhead{M($^{56}$Ni) \tablenotemark{a}}& \colhead{M$_{\text{grav,NS}}$\tablenotemark{a}} & \colhead{M$_{\text{ejecta}}$}\tablenotemark{a} & \colhead{E$_{\text{rot,NS}}$\tablenotemark{b}} &\colhead{E$_{\text{rot}}$\tablenotemark{c}} & \colhead{$|\text{E}_{\text{tot}}|$\tablenotemark{c}}
\\
\colhead{[$\mso$]}   & \colhead{[10$^{51}$ erg]} & \colhead{[$\mso$]} & \colhead{[$\mso$]} & \colhead{[$\mso$]} & \colhead{[10$^{51}$ erg]} &  \colhead{[10$^{51}$ erg]} & \colhead{[10$^{51}$ erg]}
\\
}
\setlength\extrarowheight{-1.6pt}
\startdata
4  & 1.64 & 0.12 & 1.34 & 1.83  & 0.69  & 0.0004  & 0.97 \\
5  & 2.75 & 0.22 & 1.21 & 2.76  & 0.38  & 0.0009  & 1.06 \\
6  & 1.33 & 0.12  & 1.38 & 3.35  & 4.77  & 0.0021  & 1.10 \\
7  & 0.65 & 0.07 & 1.26 & 4.3   & 1.97  & 0.0021  & 1.04 \\
8  & 0.83 & 0.09 & 1.28 & 5.07  & 2.40  & 0.0030  & 1.12 \\
9  & --   & --    & -- & --    & 21.5 &    0.0077  & 1.71 \\
10 & 4.45 & 0.11 & 1.75 & 6.07  & 24.8 &  0.0092  & 1.64 \\
11 & 2.64 & 0.13 & 1.64 & 6.98  & 20.1 &  0.0128 & 1.71 \\
12 & 1.41 & 0.09 & 1.46 & 7.98  & 8.52  & 0.0089  & 1.48 \\
13 & 2.47 & 0.15 & 1.39 & 8.82  & 5.17  & 0.0073  & 1.48 \\
14 & 3.84 & 0.21 & 1.49 & 9.47  & 6.10  & 0.0091  & 1.66 \\
15 & 4.28 & 0.20 & 1.59 & 10.1  & 13.2 &  0.0116 & 1.78 \\
16 & 5.09 & 0.09 & 1.79 & 10.6 & 30.2 &   0.0138 & 1.96 \\
17 & --   & --    & --   & --    & --    & 0.0151 & 2.08 \\
18 & --   & --    & --   & --    & --    & 0.0166 & 2.19 \\
19 & --   & --    & --   & --    & --    & 0.0183 & 2.31 \\
20 & --   & --    & --   & --    & --    & 0.0120 & 2.44 \\
21 & --   & --    & --   & --    & --    & 0.0227 & 2.56 \\
22 & --   & --    & --   & --    & --    & 0.0267 & 2.71 \\
23 & --   & --    & --   & --    & --    & 0.0267 & 2.71 \\
24 & --   & --    & --   & --    & --    & 0.0259 & 2.75 \\
25 & --   & --    & --   & --    & --    & 0.0295 & 2.79 \\
26 & 2.99 & 0.098 & 1.53 & 18.3 & 17.1  & 0.0245 & 2.61 \\
27 & 4.97 & 0.170  & 1.48 & 19.1 & 7.51  & 0.0312 & 2.86 \\
28 & 3.50 & 0.142 & 1.55 & 19.8 & 15.8  & 0.0307 & 2.87 \\
29 & 3.85 & 0.095 & 1.65 & 20.4 & 26.2  & 0.0311 & 2.95 \\
30 & 4.07 & 0.096 & 1.66 & 21.1 & 16.2 & 0.0318 & 3.04 \\
31 & 4.29 & 0.096 & 1.65 & 21.8  & 12.6 & 0.0328 & 3.12 \\
32 & --   & --    & --   & --    & --    & 0.0376 & 3.26 \\
33 & --   & --    & --   & --    & --    & 0.0389 & 3.35 \\
34 & 1.17 & 0.003 & 2.01 & 23.5  & 98.6  & 0.0421 & 3.46 \\
35 & --   & --    & --   & --    & --    & 0.0420 & 3.53 \\
36 & --   & --    & --   & --    & --    & 0.0434 & 3.64 \\
37 & --   & --    & --   & --    & --    & 0.0457 & 3.73 \\
38 & --   & --    & --   & --    & --    & 0.0479 & 3.84 \\
39 & --   & --    & --   & --    & --    & 0.0501 & 3.90 \\
40 & --   & --    & --   & --    & --    & 0.0487 & 3.99 \\
41 & --   & --    & --   & --    & --    & 0.0505 & 4.05 \\
42 & --   & --    & --   & --    & --    & 0.0556 & 4.18 \\
43 & --   & --    & --   & --    & --    & 0.0565 & 4.30 \\
44 & --   & --    & --   & --    & --    & 0.0630 & 4.38 \\
45 & --   & --    & --   & --    & --    & 0.0608 & 4.42
\enddata
\tablenotetext{a}{Resulting from applying the \cite{2016MNRAS.460..742M} model}
\tablenotetext{b}{Considering the mass of the NS from the \cite{2016MNRAS.460..742M} model, moment of inertia from \cite{2008ApJ...685..390W}, assuming conservation of angular momentum during core collapse and a NS radius of 15 km}
\tablenotetext{c}{At core collapse}

\end{deluxetable*}

\begin{figure}[h!]
\plotone{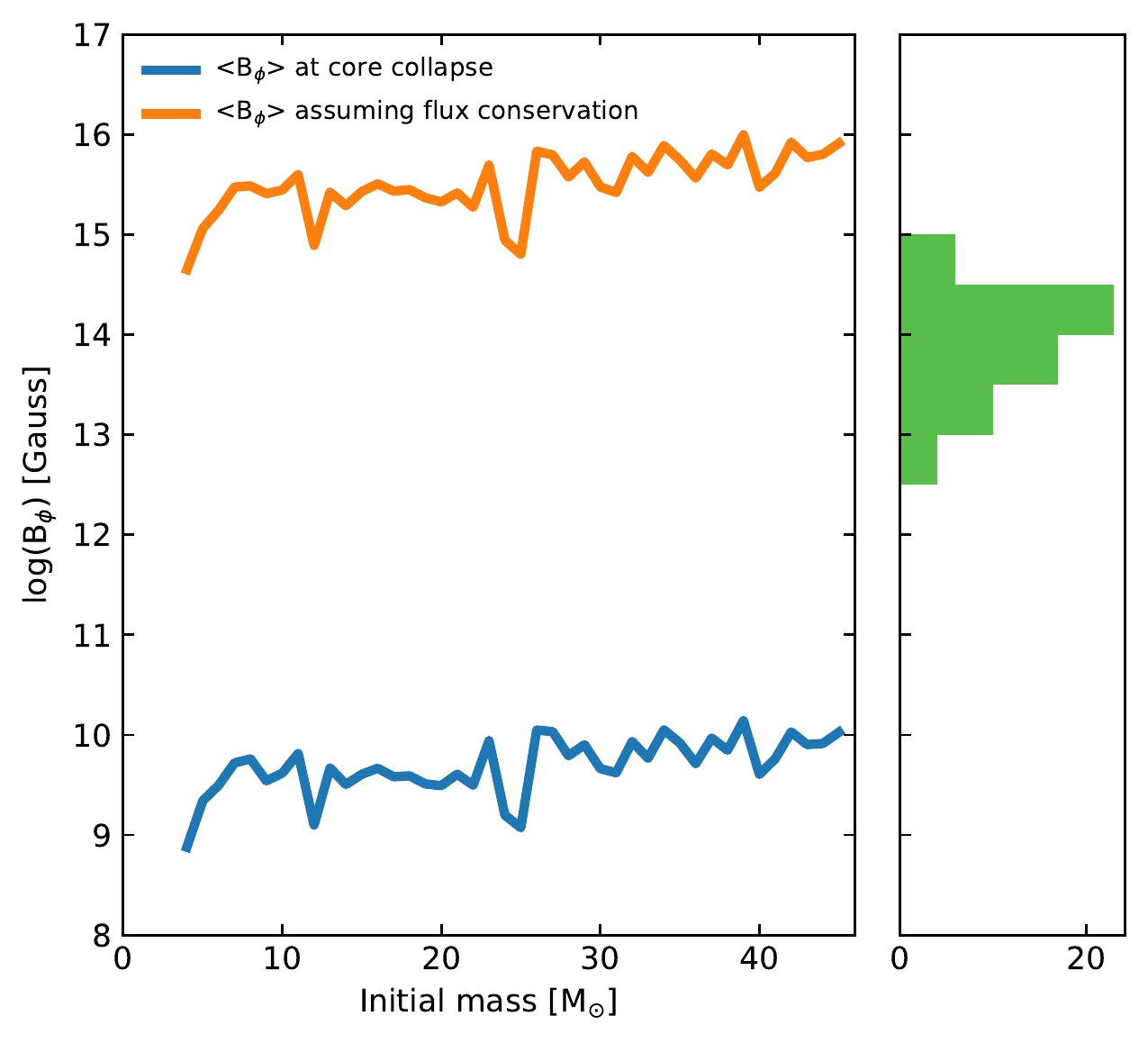}
\caption{Azimuthal component of the magnetic field, averaged over the innermost 1.5$\mso$ of our core collapse models (blue), and assuming this region contracts homologously to a radius of 15 km, and that magnetic flux is conserved during the contraction (orange), as a funciton of initial mass. Compared to the inferred values for the SLSN samples analyzed by \cite{2017ApJ...850...55N}, \cite{2018ApJ...869..166V} and \cite{2018ApJ...865....9B,2019ApJ...872...90B} in the context of the magnetar model (histogram on the right). }\label{fig:bfields}
\end{figure}

In the context of magnetorotational explosions it is also interesting to consider the initial magnetic field strength in the core.
Figure \ref{fig:bfields}, shows the strength of dynamo-generated magnetic fields averaged inside the innermost 1.5 $\mso$; namely 
\begin{equation}
\langle B_{\phi}\rangle = \frac{\int_0^{1.5 \mso}B_{\phi}(m)\text{d}m}{\int_0^{1.5 \mso}\text{d}m},
\end{equation}
at core collapse in blue, and assuming that the core contracted homologously to a radius of 15 km and that contraction was flux conserving after core collapse in orange. Magnetic fields generated by the Spruit-Tayler dynamo alone are of the order of 10$^{10}$ Gauss in the stellar cores. Despite this large strength, these fields are unlikely to be dynamically important (as opposed to rotation), since the plasma parameter $\beta$ inside the core never goes below around 10$^2$ in any of our models at core collapse (or at any other evolutionary stage).

Magnetic field strengths calculated assuming homologous contraction and flux conservation yield values that are in similar orders of magnitude than those inferred for magnetar driven SLSNe by observations \citep{2017ApJ...850...55N,2018ApJ...869..166V,2018ApJ...865....9B,2019ApJ...872...90B,2020arXiv200209508B}. However, magnetic field values at core collapse are not simply scaled to their flux-conserving value at NS radius, but are rather amplified by a dynamo during the formation of the proto-NS, either driven by convection  (e.g. \citealt{1993ApJ...408..194T,2020arXiv200306662R}), or
the magnetorotational instability (MRI, e.g., \citealt{2003ApJ...584..954A,2015ApJ...798L..22M,2015MNRAS.450.2153G,2015Natur.528..376M,2012ApJ...759..110M,2018JPhG...45h4001O}). 
These amplification mechanisms may not guarantee
fast magnetorotational explosions independent of the
seed fields. In the case of the MRI, the amplification factor may be limited \citep{2016MNRAS.460.3316R} or
it may not immediately produce strong dipole fields. Furthermore, strong dipole fields
created by dynamo action inside the proto-SN
may not affect SN dynamics on short time
scales, before they break out through the NS surface.
It is therefore noteworthy that the magnetic field strengths we predict are already very high, and little amplification is needed to match the observationally inferred values.

\section{Observable properties}\label{sec:obs}

The final fate of our evolutionary sequences will be determined to first order by whether a BH or a fast spinning magnetar is predicted to be formed by their core collapse model. BH forming fast rotators are likely to form collapsars, whereas magnetar-forming stars may produce magnetar-driven SLSNe, magnetar driven lGRBs, or perhaps both.

The method we employ to determine whether a core-collapse model results in a SLSN or a lGRB also yields an expected value for ejecta masses, nickel masses and explosion energies for the models that result in SLSNe. These are predictions that result from considering that the explosions are powered by neutrinos alone, without taking the contribution of rotation to the energy budget of the explosion. However, taking these results at face value implies that, even without the extra energy, these explosions would anyway have longer lifetimes and slower rise times than typical stripped-envelope SNe. This follows from the simple approach of applying the model of \cite{1982ApJ...253..785A}. However, the explosion energies and nickel masses alone cannot explain the high luminosities of SLSNe, and the additional energy source is still required.

\begin{figure}[h!]
\plotone{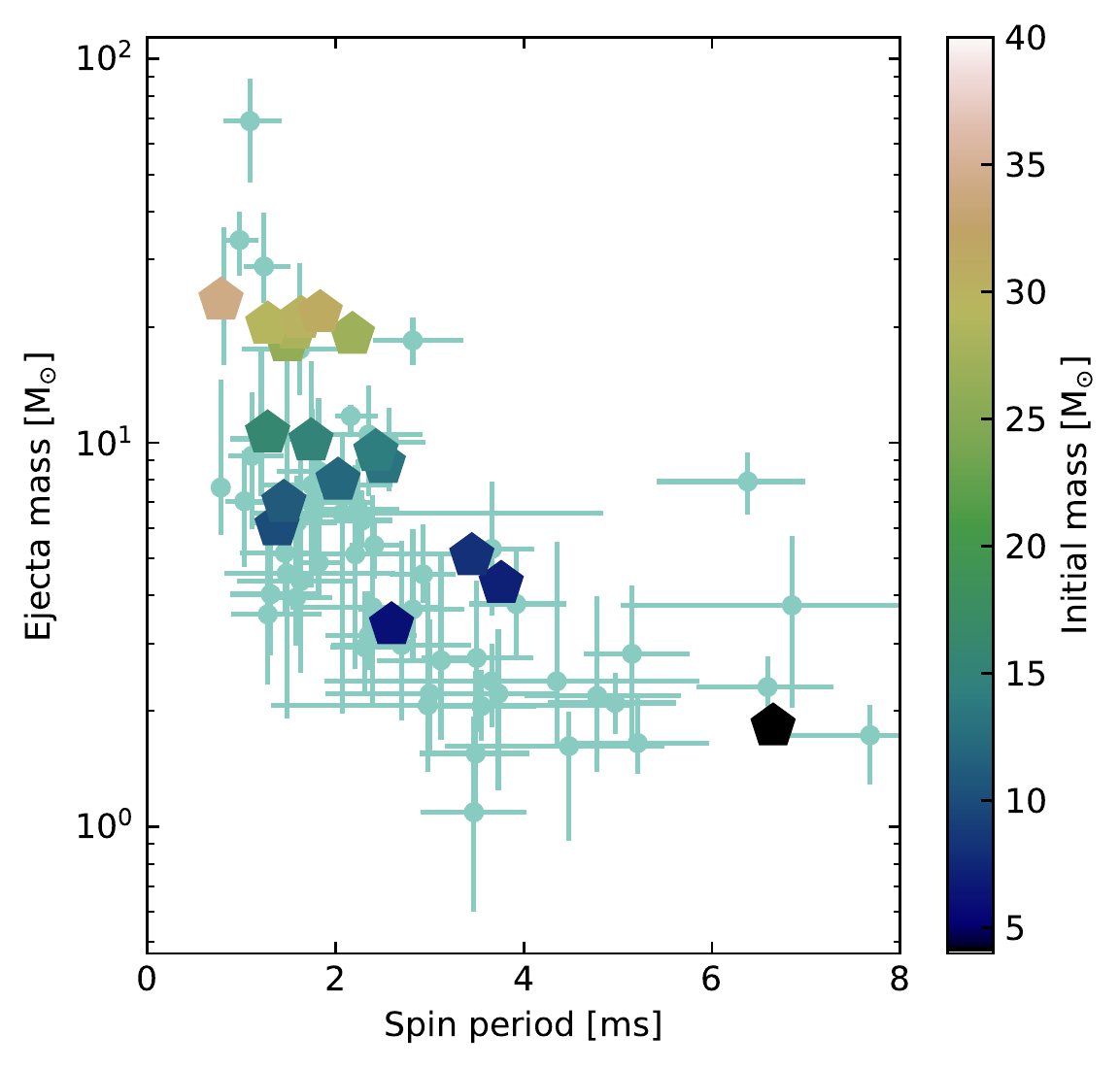}
\caption{Pentagons represent the ejecta masses produced by SN explosions produced from the our core collapse models --as predicted by the \cite{2016MNRAS.460..742M} model-- as a function of the spin periods calculated for their remnants, and colored by their initial mass. Spin periods are calculated assuming homologous contraction of the collapsing core to a radius of 15 km, conservation of angular momentum, and NS moment of inertia as prescribed by \cite{2008ApJ...685..390W}. Light blue points correspond to the inferred values from observations of SLSNe obtained by  \cite{2017ApJ...850...55N}, \cite{2018ApJ...869..166V} and \cite{2018ApJ...865....9B,2019ApJ...872...90B}.}\label{fig:spin}
\end{figure}

Assuming that our evolutionary sequences result in magnetar formation yields consistent results with the observed properties of SLSNe inferred by the samples analyzed by \cite{2017ApJ...850...55N}, \cite{2018ApJ...869..166V} and \cite{2020arXiv200209508B}. Figure \ref{fig:spin} shows the expected ejecta masses (taken from Table \ref{tab:table2}) and spin periods expected from the evolutionary sequences that explode as SLSNe. The spin period is calculated assuming that the NS formed after collapse contains the angular momentum remaining on the mass that is not ejected, and that angular momentum is conserved and reconfigured into a radius of 15 km. The moment of inertia of the newly-formed NS is taken from \cite{2008ApJ...685..390W}. These values seem to be consistent with those inferred from observations, and match correlation between spin period and ejecta mass found by \cite{2020arXiv200209508B}. In our models, this correlation is caused to first order by the different life times that stars have as a function of initial mass: Less massive stars live longer, and thus have more time to transport angular momentum outside of their contracting iron core. 

\begin{figure}[h!]
\plotone{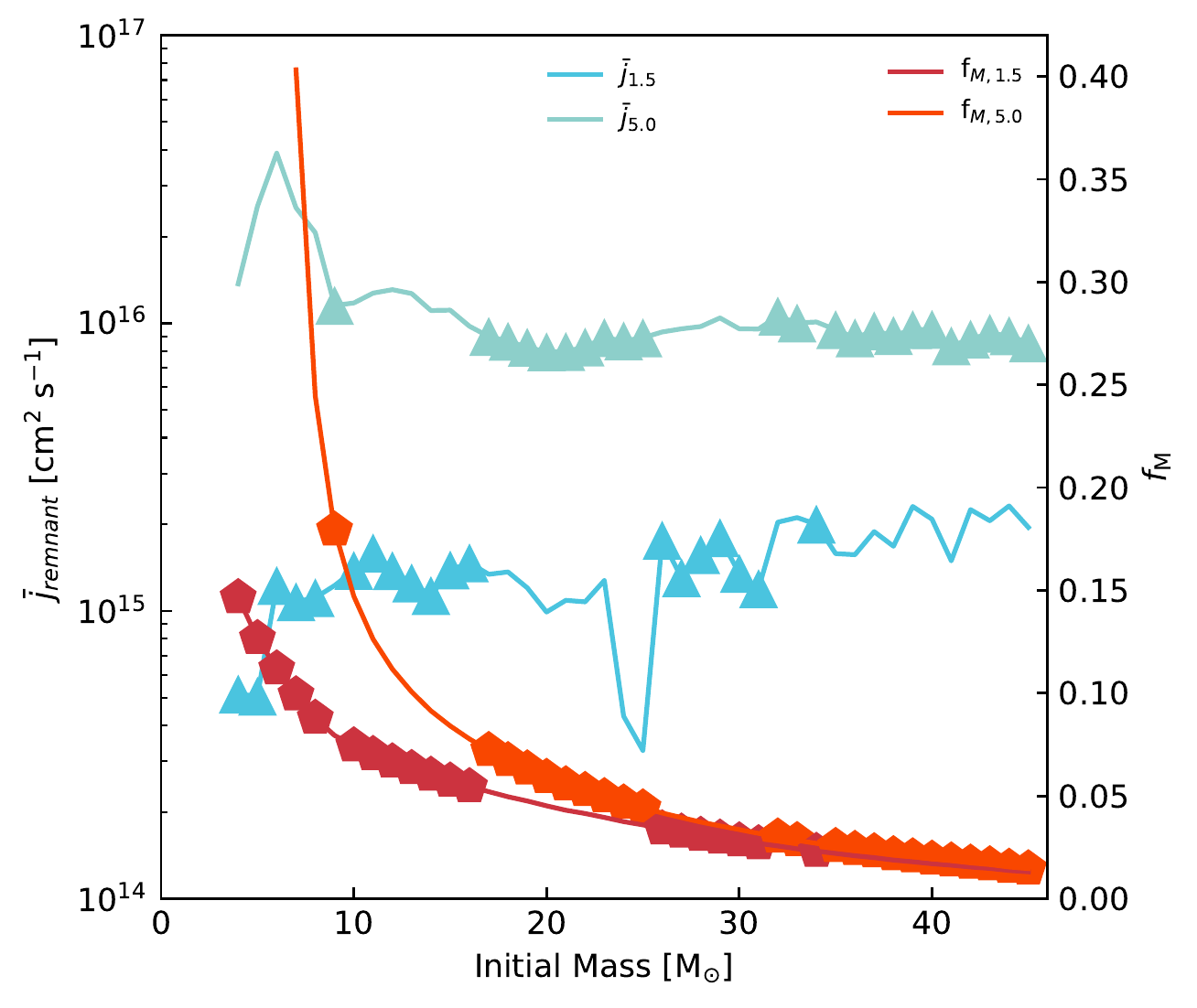}
\caption{The left y-axis shows the average specific angular momentum in the innermost 1.5 $\mso$ (blue) and 5 $\mso$ (light green) of core collapse models, chosen as mass coordinates representative of NS and collapsar forming models, respectively. On the right y-axis, the fraction of the kinetic energy that can be converted into radiation by CSM interaction, $f_M$, is shown. This is calculated assuming that the explosions leave behind a 1.5 $\mso$ remnant (red) and a 5 $\mso$ remnant (orange), and that the ejected mass interacts with the CSM mass given by $\Delta \text{M}_{\text{He} \rightarrow \text{final}}$. Hexagons in the plot correspond to core collapse models that explode as a SLSN according to the \cite{2016MNRAS.460..742M} model, triangles correspond to models that are expected to form a BH. All quantities shown as a function of initial mass.}\label{fig:jandm}
\end{figure}

Another feature that will become important, particularly for lower mass progenitors, is how the ejecta will interact with the CSM. Even with lower mass loss rates than in ALMS18, it is expected that the ejecta will encounter a massive CSM in a non-trivial geometry, given the fact that the mass loss rates increase steeply during the last 1000 years before core collapse, and a non-spherical structure of around 0.5 $\mso$ will be located in a likely toroidal configuration around the star. Determining the exact location and density profile of this structure requires, at least, 2D numerical simulations of the flow, which will be concentrated in the equatorial plane, and which will likely have low velocities, since it is accelerated by the centrifugal force of material at the surface, and not by the radiation field of the star, as is the case in a Wolf-Rayet wind (see \cite{2008A&A...478..769V} for simulations of similar structures, and ALMS18 for a detailed discussion). However, as shown with the red and orange lines in Figure \ref{fig:jandm}, the mass contained in this region can convert a fraction $f_M$ of the kinetic energy of the ejecta into radiation, that goes up to 15\% in the lowest mass models. This could possibly help in explaining bumps in SLSNe light curves, or possibly becoming important several days after maximum, depending on the geometry of the CSM (see Figure 12 of ALMS18).

Figure \ref{fig:jandm} also shows average specific angular momentum in the innermost 1.5 $\mso$, corresponding to a case that forms a NS, and 5 $\mso$, taken as a fiducial mass for a BH, indicating which progenitor models we expect to successfully explode, and which ones we expect to form a BH with a triangle in models with $\xi_{2.5}$ larger or smaller than 4.5. Those that we expect to form BHs are going to have enough angular momentum to form collapsars, whereas those that form NSs have angular momenta consistent with magnetars (also seen in Figure \ref{fig:spin}).

\section{Conclusions}\label{sec:conclusions}

We computed a grid of evolutionary models for fast rotating, low metallicity stars with enhanced rotational mixing, and found that changing the way that angular momentum is lost in response to mass loss changed the expected outcome with respect to ALMS18. We found larger final masses, a higher helium content and less massive CSM. However, we still expect them to be viable progenitors for Type\,Ic SLSNe and lGRBs, and we expect that interaction with the nearby CSM will manifest itself in the lightcurve in an observable way, particularly in models with low ejecta mass, which will be closer to the expected CSM mass.

We studied the explodability of our core collapse models through their compactness parameter $\xi_{2.5}$ \citep{2011ApJ...730...70O}, the parameters $M_4$ and $\mu_4$ \citep{2016ApJ...818..124E}, and an analytical test proposed by \cite{2016MNRAS.460..742M}. These three tests have been used to estimate the likelihood of non-rotating stellar evolutionary sequences during core collapse to form either a NS --and a successful SN explosion-- or formation of a BH, in the neutrino driven SN scenario. Given their angular momenta and magnetic field strengths, these progenitors might correspond to progenitors of either lGRBs or SLSNe, depending on the result of the collapse. We find that the behavior of these tests is similar to non-rotating stars, presenting a non-linear behavior as a function of initial mass, but rotational mixing can affect the exact evolution of the stellar core by changing the relation between initial mass and subsequent core masses, and also smoothing composition gradients throughout the evolution, therefore affecting the explodability of these stars.

We found that, taking $\xi_{2.5}$ as an indicator of the likelihood of forming either NSs or BHs during core collapse, it is possible that the initial mass is not enough to determine the fate of a quasi-chemically homogeneous massive star. This may result in either the explosion forming a SLSN or a lGRB, depending on whether a magnetar or a BH is formed.

However, a significant amount of rotational energy can be released by the contracting proto-NS during collapse, and the explosion should also be promoted by rotationally and magnetically driven instabilities, which might relax the value of the compactness parameter that separates exploding and imploding cores. \added{(see \citealt{2018ApJ...852...28S} for the beneficial influence of rapid rotation in the explosion mechanism).} Nevertheless, only detailed simulations of the explosion could give a definite answer.
Regardless of the initial mass of the progenitor, our core collapse models resemble the observed ejecta masses and rotation periods inferred for SLSNe, if indeed magnetars are the main source of energy during the explosion, even when considering more efficient angular momentum loss.

We also found that fast rotating stars at low metallicity are consistently strongly magnetic in their cores, which might lead to the magnetic fields being important during the process of collapse, although not dynamically during the evolution of the star.

We found that the evolutionary channel experienced by the evolutionary sequences presented in this work will produce lGRBs and SLSNe, consistent with the inferred properties of their progenitors, and likely as a non-monotonic function of initial mass. In particular, our models reproduce the range of ejecta masses, magnetar spin periods and magnetic fields inferred for SLSNe \citep{2020arXiv200209508B}. A more detailed study of this channel in different metallicities and with different rotational velocities is required, however, to compare with the SLSN and lGRB rates, and with the ejecta mass distribution inferred by \cite{2020arXiv200209508B}. These transients might have important CSM interaction, particularly if observed close to their plane of rotation.

\acknowledgments
Acknowledgments. 
DRAD would like to acknowledge valuable discussions with Matteo Bugli, Luc Dessart, Eva Laplace, Abel Schootemeijer and Alejandro Vigna-Gomez.
BM was supported by the Australian Research Council (ARC) through
Future Fellowship FT160100035
and as an associate investigator of the ARC CoE for Gravitational Wave Discovery \emph{OzGrav} (CE170100004).
 
\software{MESA \citep{MESAI,MESAII,MESAIII,2018ApJS..234...34P}, \texttt{matplotlib} \citep{Hunter:2007},\texttt{numpy} \citep{2011CSE....13b..22V}}

\bibliographystyle{aasjournal}
\bibliography{references}

\end{document}